\documentclass [10pt]{iopart}
\usepackage{epsf}
\eqnobysec
\makeatletter
\long\def\@makefntext#1{\parindent 1em\noindent
 \makebox[1em][l]{\footnotesize\rm$\m@th{{}^\arabic{footnote}}$}%
 \footnotesize\rm #1}
\def\@makefnmark{${}^\arabic{footnote}$}
\def\@thefnmark{${}^\arabic{footnote}$}
\makeatother
\begin{document}
\title{Static fluid cylinders and their fields: global solutions}
\author{J. Bi\v{c}\'{a}k\dag\ddag, T. Ledvinka\dag, B. G. Schmidt\ddag and M. \v{Z}ofka\dag}
\address{\dag Institute of Theoretical Physics, Faculty of Mathematics and Physics, Charles University Prague, V Hole\v{s}ovi\v{c}k\'{a}ch 2, 180 00 Praha 8, Czech Republic\\
\ddag Albert-Einstein Institute, Max-Planck-Institute for Gravitational Physics, Golm, Germany\\}
\eads{\mailto{bicak@mbox.troja.mff.cuni.cz}\\\mailto{ledvinka@mbox.troja.mff.cuni.cz}\\
\mailto{bernd@aei-potsdam.mpg.de}\\ \mailto{zofka@mbox.troja.mff.cuni.cz}}
\date{}
\begin{abstract}
The global properties of static perfect-fluid cylinders and their external Levi-Civita fields are studied both analytically and numerically. The existence and uniqueness of global solutions is demonstrated for a fairly general equation of state of the fluid. In the case of a fluid admitting a non-vanishing density for zero pressure, it is shown that the cylinder's radius has to be finite. For incompressible fluid, the field equations are solved analytically for nearly Newtonian cylinders and numerically in fully relativistic situations. Various physical quantities such as proper and circumferential radii, external conicity parameter and masses per unit proper/coordinate length are exhibited graphically.
\end{abstract}
\submitto{\CQG}
\section{Introduction and summary}
Since 1917 when Levi-Civita gave his static vacuum solution, cylindrically symmetric spacetimes have played an important role in general relativity. Recently, attention has been mainly paid to dynamical situations or quantum issues. Some of these aspects are surveyed in \cite{Bicak} (section 9 and references [247-263], [275-277] therein), for example; for even more recent works, see, e.g., \cite{Bondi, Mena, Goncalves}.

Still, in a recent concise review on static cylinders \cite{Bonnor}, Bonnor introduces the topic by stating that `the Levi-Civita spacetime continues to puzzle relativists'. In contrast to the Schwarzschild metric described completely by a single parameter, the Levi-Civita solution contains two essential constant parameters---$m$, related to the local curvature, and $\mathcal{C}$, determining the conicity of the spacetime. Relating these two parameters to physical sources turns out to be delicate, as we recently found in the case of static cylindrical shells of various types of matter \cite{BZ}, and as we will demonstrate for solid perfect-fluid cylinders in the present paper.

In the second edition of the `exact-solutions bible' \cite{Exact Solutions}, there are quoted about 20 papers on static perfect-fluid cylindrically symmetric fields. However, these solutions are local, no analysis of global properties is usually available. In addition, most perfect-fluid equations of state in the literature are either {\it ad hoc} or not very physically plausible. In fact, the cylindrical counterpart of the famous Schwarzschild interior solution, in which the fluid is incompressible, so that its matter density $\mu = \mu_0 =$ constant, is not known. Although this equation of state is unphysical (implying infinite velocity of sound), in the spherical case not only can the corresponding solution be easily found but it also gives `not unreasonable' estimates for maximum masses of neutron stars, for example.

The purpose of the present work is to study static perfect-fluid cylinders and their fields globally, to investigate the Newtonian limit and to analyse incompressible cylinders numerically. In section 2, we start from a line element in which the 3-space is rescaled by the norm of the timelike Killing vector and we choose the distance from the axis as the radial coordinate. Usually we require the axis to be regular, but the presence of a conical singularity (a `cosmic string') along the axis is also considered. The field equations in these coordinates enable one to find convenient relations between some of the metric functions. In particular, the vacuum Levi-Civita solution can be rederived in these coordinates. We show in detail that although it contains five constants of integration, only two---the mass parameter and the conicity parameter---cannot be removed.

In section 3 we first formulate a lemma (and then prove it in appendix A) stating that for a smooth equation of state $\mu(p)$ and any value of the central density $\mu_0$ there exists a unique solution of the field equations in a neighbourhood of the axis, which can be continued into a global solution. Depending on the equation of state, the fluid occupies the entire space, or the cylinder is of a finite extent in the radial direction provided that the pressure vanishes at a finite radius, $p(R)=0$. Another lemma shows that a unique vacuum solution can be joined to the inner perfect-fluid solution at this radius $R$ where the pressure vanishes. Finally, a Theorem is proven asserting that systems with $\mu(p=0)>0$ always have a finite radius---like with spherically symmetric balls.

The field equations are written in terms of $\lambda = 1/c^2$ ($c$---the velocity of light); for $\lambda=0$ the Newtonian limit is recovered. The conformal 3-space becomes flat and the relativistic Tolman mass turns into the Newtonian mass per unit length. This is demonstrated in section 4. In appendix B, the Newtonian static constant-density cylinders and spheres are constructed directly within the Newtonian theory for comparison. (For shell sources in Newtonian gravity, see appendix in \cite{BZ}.)

In the brief section 5 it is explicitly shown how an outer Levi-Civita solution can be joined to an inner perfect-fluid solution at the boundary where the pressure vanishes. The need of a non-trivial conicity parameter, $\mathcal{C} \not =1$, outside a general perfect-fluid cylinder is discussed in detail in appendix C. As an example, the Evans solution \cite{Evans} with the equation of state $\mu = \mu_0 +5p$, $\mu_0 >0$, is employed to elucidate that only special cylinders can be joined to the outer Levi-Civita metrics with the conicity parameter $\mathcal{C}=1$.

Section 6 is devoted to cylinders of incompressible fluid. As mentioned above, exact solutions representing incompressible perfect fluid cylinders have not yet been found. We first study analytically weakly gravitating cylinders. Expanding all functions entering the field equations in terms of the dimensionless radial distance from the axis, the pressure at any point is found in terms of the central pressure $p_c$. Inverting the series, the radius of a cylinder can be determined by requiring the pressure to vanish. In this way, relativistic parameters characterizing the cylinder can be determined analytically as corrections to the Newtonian values. The conicity parameter starts to deviate from its Minkowskian value only by terms of the order $O(\Pi_c^2 \ln \Pi_c)$ where $\Pi_c = p_c/\mu_0 c^2$.

Fully relativistic cylinders of incompressible fluid are treated numerically. Here we extend considerably the results of Stela and Kramer \cite{Stela and Kramer}. We allow the conicity parameter $\mathcal{C} \not =1$ (as one should---see above), admit high values of the central pressure and we also exhibit other quantities of interest graphically. In particular, we show how the proper radius of the cylinder increases with increasing central pressure, whereas the circumferential radius starts to decrease. The peculiar behaviour of the circumferential radius is well seen in the embedding diagram of the $(z,t)=$ constant surfaces and their dependence on the central pressure. Both the external Levi-Civita mass parameter and external conicity parameter are plotted as functions of the central pressure. It is illustrated graphically that the dimensionless mass per unit proper length has an upper bound of $1/4$, whereas Thorne's C-energy scalar is bounded from above by $1/8$. We also plot dimensionless mass per unit coordinate/proper length of the cylinders as functions of their proper and circumferential radii and compare them with the case of incompressible spherical stars.

We believe that full understanding of static cylindrical systems will give intuition for more realistic situations in the neighbourhood of elongated but finite systems. In addition, the only realistic hope to construct exact examples of interaction between simple physical matter and gravitational waves seems to live in cylindrical symmetry. Even perturbative calculations of slowly rotating and oscillating cylinders might bring useful insights. The simple static configurations would then serve as convenient backgrounds.
\section{Field equations and their vacuum solutions}
We define cylindrical symmetry for static non-vacuum spacetimes by assuming that in addition to the timelike Killing vector, two further commuting Killing vectors acting in hypersurfaces orthogonal to the `static' Killing vector exist\footnote[1]{For the definition of cylindrical symmetry, see \cite{Exact Solutions}; for a detailed, careful discussion, see, e.g., the recent work \cite{Carot et al.}.}. Adapting coordinates to the symmetries, we can write the line element with the metric in the 3-space rescaled by the norm of the timelike Killing vector (written as an exponential) in the form
\begin{equation}\label{the original metric}
ds^2 = -e^{2U\over c^2}c^2dt^2+e^{-{2U\over c^2}}(A^2dr^2 +B^2 d\varphi^2
+C^2dz^2)\ ,
\end{equation}
with $ A(r),U(r), B(r),C(r) $, where we keep the velocity of
light $c$ because we wish to discuss the Newtonian limit in detail. We assume
circular orbits for ${\partial\over \partial\varphi}$ and consider only spacetimes with a regular axis. Occasionally, we admit an infinitely thin cosmic string, which is described by a conical singularity along the axis. The distance from the axis (in the conformal 3-metric) can be used to define a unique radial coordinate, i.e., we put $A=1$. There are
many other geometrical choices for the radial coordinate. If we fix
the range of $\varphi$ to be $[0,2\pi)$, we obtain the following
regularity conditions on the axis by comparing our form of the metric with the Minkowskian metric (possibly containing the conical singularity) written in cylindrical coordinates:
\begin{equation}\label{Initial conditions}
\begin{array}{rclrcl}
  U(0) & = & 0, \hspace{1cm} U\:'(0) & = & 0,\\
  B(0) & = & 0, \hspace{1cm} B\:'(0) & = & 1/\mathcal{C}^*,\\
  C(0) & = & 1, \hspace{1cm} C\:'(0) & = & 0,
\end{array}
\end{equation}
where $\mathcal{C}^*$ is the axis-conicity; in particular, the axis is regular for $\mathcal{C}^* = 1$. The functions $U, r^{-1}B$ and $C$ have to be differentiable
functions of $r^2$ to ensure differentiability on the axis. The
values (\ref{Initial conditions}) of $U$ and $C$ on the axis determine the normalization of the static and translational Killing vectors.

For fluid cylinders we have
\begin{equation}
T_{\alpha\beta}=(c^2 \mu+p)u_\alpha u_\beta+pg_{\alpha\beta}\ ,
\end{equation}
where the 4-velocity $u^\alpha$ is normalized as
\begin{equation}
 g_{\alpha\beta}u^\alpha u^\beta=-1.
\end{equation}
The fluid is assumed to be at rest in the coordinates of metric (\ref{the original metric}), hence $u^\alpha = ( c^{-1} \exp (-U/c^2),0,0,0)$. The Einstein field equations
\begin{equation}
R_{\alpha\beta}={8\pi G\over c^4}(T_{\alpha\beta}-{1\over 2}T g_{\alpha\beta})
\Longleftrightarrow G_{\alpha\beta}={8\pi G\over c^4}T_{\alpha\beta}\ ,
\end{equation}
imply ($\lambda={1\over c^2}$)\footnote{To avoid confusion, hereafter we denote the inverse square of the velocity of light by $\lambda$. Since $C$ is a metric function, we denote the `external' Levi-Civita conicity parameter by $\mathcal{C}$ (cf. equation \eref{Transformation} and below) and the `internal' axis-conicity by $\mathcal{C}^*$ as in equation \eref{Initial conditions}.}:
\begin{equation}
R_{tt}=U''BC +B'CU'+C'BU'=4\pi G (\mu+\lambda3p)e^{-2U\lambda}BC\
, \label{equationTT}
\end{equation}
\begin{equation}
G_{rr}=-\lambda^2(U')^2 BC +B'C'=8\pi G e^{-2U\lambda}BC\lambda^2
p\ , \label{equationRR}
\end{equation}
\begin{equation}
G_{\varphi\varphi}=\lambda^2(U')^2 C+C''=8\pi G e^{-2U\lambda}
C\lambda^2p\ , \label{equationFF}
\end{equation}
\begin{equation}
G_{zz}=\lambda^2(U')^2 B+B''=8\pi G e^{-2U\lambda} B\lambda^2p\ .
\label{equationZZ}
\end{equation}
As we have three unknown functions and three second order equations \eref{equationTT}, \eref{equationFF}, \eref{equationZZ} we
expect six integration constants. The first-order equation \eref{equationRR} reduces the number of constants in the solution to five. The conservation law $T^\alpha{}_{\beta;\alpha}$=0 gives
\begin{equation}\label{U of p}
U'=-{p'\over \lambda p+\mu}\ .
\end{equation}
We assume that an equation of state (EOS) $ \mu(p)$ is given.
Then we can integrate the last equation,
\begin{equation}
U(r)-U(0)=U(r) =-\int_{p_c}^{p(r)}{dp\over \lambda p+\mu(p)}
\label{Integration of U}\ .
\end{equation}
For incompressible fluid of density $\mu_0$ and central pressure $p_c$, this can be integrated explicitly to yield
\begin{equation}\label{Integrated U}
 U(R) = \frac {1} {\lambda } \ln (1 + \lambda \frac {p_c} {\mu_0})
\end{equation}
for the value of $U$ on the surface $r=R$ with $p(R)=0$. The field equations can be simplified as follows: from equations (\ref{equationRR}), (\ref{equationFF}) and (\ref{equationZZ}) we can eliminate $U'$ to obtain
\begin{equation}
{(BC)''\over BC}=8\pi G\lambda^2 e^{-2U} 4p \label{Only BC
Equation} .
\end{equation}
\Eref{equationTT} can be rewritten as
\begin{equation}
U'' +{(BC)'\over BC}U'=4\pi G (\mu+\lambda 3p)e^{-2U\lambda}
\label{Only BC and U Equation} .
\end{equation}
Hence, we obtain two equations for $U$ and for the product $BC$.

One immediate consequence of the field equations: take $C \times$ (\ref{equationZZ})$-B \times$ (\ref{equationFF}) and consider the above initial conditions
(\ref{Initial conditions}). This yields
\begin{equation}\label{Constraint}
CB\:'-BC\:'= \frac{1}{\mathcal{C}^*} = \mbox{constant}
\end{equation}
for $0 \leq r < \infty$.

To construct local solutions having a prescribed EOS and ignoring the regularity of the axis, one can proceed as follows: given $\mu(p)$, equation (\ref{Integration of U}) implies there are unique functions $\mu(U)$, $p(U)$. Inserting these functions in equations (\ref{Only BC Equation}), (\ref{Only BC and U Equation}), we can solve for $U$ and $BC$ locally. Next, equations (\ref{equationFF}) and (\ref{equationZZ}) determine $B$ and $C$. \Eref{equationRR} is satisfied as a consequence of the Bianchi identities. Hence, given an EOS, there are
many local solutions away from the axis. Infinitely extended solutions with some special equations of state ($\mu = \gamma p$) were derived in \cite{Kramer-full, Teixeira, Bronnikov}, while finite solutions satisfying energy conditions were obtained in \cite{Haggag} and \cite{Evans}. In \cite{Scheel}, the authors consider polytropic equations of state using numerical solutions of the structure equations.

We can also easily find {\it vacuum} solutions in the above coordinates. For simplicity, we now put $\lambda = 1/c^2 = 1$. \Eref{Only BC Equation} implies $BC = a_1(r+a_2)$ with constant parameters $a_1 \not =0$ and $a_2$  (if $a_1=0$ then either $B=0$ or $C=0$ in contradiction with (\ref{the original metric})). From (\ref{equationTT}), we have $(U'BC)'=0$, so we can write $U'BC = a_1a_3$ where we introduced a third integration constant $a_3$\footnote{Note its relation to the Tolman mass, $m_T$ of \eref{Definition of Tolman Mass}, following from $U'BC = 2m_T =$ constant.}. Substituting for $BC$ we find
\begin{equation}\label{U}
U= a_3\ln \left( {r+a_2\over a_4} \right),
\end{equation}
with $a_4 \not =0$ being a fourth integration constant. To find $B$ and $C$, we have to solve (\ref{equationFF}) and (\ref{equationZZ}) with vanishing rhs, taking into account $BC = a_1(r+a_2)$. The solutions can be written as
\begin{equation}\label{Original B,C}
B = a_1 a_5 \left( {r+a_2 \over a_5} \right)^{n_2} \hspace{-0.3cm}, \hspace{1cm} C = \left( {r+a_2 \over a_5}\right)^{n_1} \hspace{-0.3cm},
\end{equation}
where $a_5 \not =0$ is a fifth integration constant and $n_{1,2} = (1 \pm \sqrt {1-4a_3^2})/2$. (We can also interchange the roles of $B$ and $C$. As long as we have no axis we cannot really distinguish between the two Killing vectors $\partial_z$ and $\partial_\varphi$.) From the relation for $n_{1,2}$ we can see that the range of $a_3$ for which solutions exist is $a_3 \in [-1/2,1/2]$. It is thus advantageous to introduce a new constant $\gamma \in [0,1]$ such that $a_3 = \sqrt {\gamma (1-\gamma)}$. With this definition, the exponents in (\ref{Original B,C}) can be written $n_1 = \gamma, n_2 = 1 - \gamma$. Introducing $\delta = a_1 a_5$, $a = a_5 / a_4$, $b=a_2$ and $L=a_5$, the resulting metric functions read
\begin{equation}\label{LC with 5 constants}
\fl U  = \sqrt { \gamma (1-\gamma)}\ln \left( a {r+b\over L} \right), \hspace{0.7cm} C = \left( {r+b\over L} \right)^\gamma, \hspace{0.7cm} B = \delta \left( {r+b\over L} \right)^{1-\gamma}.
\end{equation}
The solution thus has five constants of integration. However, we now show that three of these constants are redundant. We shall see that this is related to the fact that the Killing vectors $\partial_t, \partial_z$ are defined only up to constant multiplicative factors and, in addition, the whole metric can be rescaled by a constant. First, denote $\alpha = \left[ a (1-\gamma) \right]^{\sqrt{\gamma (1-\gamma)}}$ and introduce new constants $\rho_0, m$ and $\tilde{\mathcal{C}}$ by the relations
\begin{equation}\label{Transformation}
L = \alpha \rho_0, \hspace{0.3cm} \gamma = \frac {m^2} {1+m^2}, \hspace{0.3cm} \delta = \frac {\alpha \rho_0} {\tilde{\mathcal{C}}} (1+m^2)^\frac {1}{1+m^2}.
\end{equation}
Then apply the transformation
\begin{equation}
[t, z, r, \varphi] \rightarrow [{\tau \over \alpha}, \: \alpha (1+m^2)^\frac {m^2} {1+m^2} \zeta, \: \frac {\alpha \rho_0} {1+m^2} \left( \frac {\rho} {\rho_0} \right)^{1+m^2} \hspace{-0.8cm} \;\; - b, \: \varphi].
\end{equation}
This leads to the metric of the form
\begin{equation}\label{LC with 3 constants}
\fl \hspace{1cm} ds^2 = - \left (\frac {\rho} {\rho_0} \right) ^{2m} \hspace{-0.3cm} d\tau^2 +
\left( \frac {\rho} {\rho_0} \right)^{2m(m-1)} \hspace{-0.3cm} (d\zeta^2 +
d\rho^2) + \frac {1}{\tilde{\mathcal{C}}^2} \rho^2 \left( \frac {\rho}
{\rho_0} \right)^{-2m} \hspace{-0.3cm} d\varphi^2,
\end{equation}
containing three constants only.

Next, use a linear coordinate transformation with a scaling parameter $q$, which will be fixed subsequently, as follows:
\begin{equation}\label{LC-LC'}
[\tau, \zeta, \rho, \varphi] \rightarrow [q^\frac {m} {m^2-m+1} t, \: q^\frac {m(m-1)} {m^2-m+1} z, \: q^\frac {m(m-1)} {m^2-m+1} r, \: \varphi].
\end{equation}
This yields
\begin{equation}\label{transformed LC}
\fl \hspace{0.3cm} ds^2 = - \left( \frac {rq} {\rho_0} \right) ^{2m} \hspace{-0.3cm} dt^2 + \left( \frac {rq} {\rho_0} \right)^{2m(m-1)} \hspace{-0.7cm} (dz^2 +
dr^2) + \left( \frac {1} {\tilde{\mathcal{C}}} \: q^\frac {m^2} {m^2-m+1} \right)^2 \left( \frac {rq} {\rho_0} \right)^{-2m} \hspace{-0.3cm} r^2 d\varphi^2.
\end{equation}
By choosing $q$ appropriately, we can dispose of either $\tilde{\mathcal{C}}$ or $\rho_0$ so that the resulting metric involves only two constants. Indeed, setting $q=\rho_0$ we eliminate $\rho_0$ in (\ref{transformed LC}):
\begin{equation}\label{LC without rho_0}
\fl \hspace{1cm} ds^2 = - r^{2m} dt^2 +
r^{2m(m-1)} (dz^2 +
dr^2) + \frac {1}{\mathcal{C}^2} r^{2(1-m)} d\varphi^2,
\end{equation}
where we put $\mathcal{C} = \tilde{\mathcal{C}}/\rho_0^\frac {m^2} {m^2-m+1}$. With $\mathcal{C} = 1$, this is the standard Levi-Civita metric as given in the primary reference \cite{Exact Solutions}, equation (22.7). In (\ref{LC without rho_0}), the $t,z$ and $r$ coordinates and the conicity parameter $\mathcal{C}$ do not have the usual dimensions. We can maintain the conventional dimensions if we set $q=\tilde{\mathcal{C}}^\frac {m^2-m+1} {m^2}$, thus obtaining
\begin{equation}\label{LC without C}
\fl \hspace{1cm} ds^2 = - \left (\frac {r} {R} \right) ^{2m} dt^2 +
\left( \frac {r} {R} \right)^{2m(m-1)} (dz^2 +
dr^2) + r^2 \left( \frac {r}
{R} \right)^{-2m} d\varphi^2,
\end{equation}
with $R=\rho_0/\tilde{\mathcal{C}}^\frac {m^2-m+1} {m^2}$. Notice, however, that this cannot be done for $m=0$. Hence, this coordinate system has the usual dimensions but (\ref{LC without C}) does not include the standard cosmic string as the limit $m \rightarrow 0$ only yields the flat spacetime without deficit angle. In the following, we use the form (\ref{LC without rho_0}).

Note that if the $\partial_\varphi$ symmetry is considered locally only, we can also rescale $\varphi$ and thus put $\mathcal{C}=1$ in (\ref{LC without rho_0}). This is not the case, however, as we insist on a particular range of $\varphi$ and require $\varphi \in [0,2\pi)$.
\section{Existence of global solutions}\label{Global Solutions}
Our aim here is to show that for a given barotropic EOS, a unique global solution corresponds to each value of the pressure on the axis $p_c>0$\footnote{We choose the pressure rather than the energy density as the fundamental variable since in section \ref{Numerics} we analyze cylinders of incompressible fluids.}. Depending on the EOS, the fluid cylinder either is of a finite
extent, or the fluid occupies the entire space. (In this section we again put $\lambda = 1/c^2 = 1$.)

We begin with the construction of solutions with a regular axis.
Inserting Taylor series for the metric coefficients into the
field equations indicates that for a given analytic equation of
state $\mu(p)$, a value $p_c$ of the pressure at the center
determines a unique solution. We prove the following lemma in \ref{Regular Axis}.

\vspace{0.3cm}

\noindent {\bf Lemma 1:} {\it Let $\mu(p)$ be a barotropic equation of state. Then for each value of the central density $\mu_0=\mu(p_c)>0$ there is $r_0>0$ such that a unique solution of the field equations (\ref{equationTT})--(\ref{U of p}) exists for $0\leq r\leq r_0$.}

\vspace{0.3cm}

Next we show that the solution can be extended as long as the pressure is positive.

\vspace{0.3cm}

\noindent {\bf Lemma 2:} {\it Suppose we have a solution with a regular axis. Then $U$, $B$ and $C$ are monotonically increasing and $p$ is monotonically decreasing as long as $p>0$ (then also $\mu + p>0$).}

\vspace{0.3cm}

\noindent {\bf Proof:} $B'$ and $C'$ are positive near the center. If any of them vanished at some point, then equation \eref{equationRR} would give a contradiction because we
assumed $p>0$, and $B,C>0$ near the center as a consequence of the
regularity conditions (\ref{Initial conditions}). Hence, $B$ and $C$ are monotonically
increasing.

\Eref{Only BC and U Equation} implies---due to $U'(0)=0$---that
$U''(0)>0$. Hence $U'$ is positive near the center. Assume
$U'(r_0)=0$; then $U''(r_0)>0$ means that $U$ cannot have a maximum at
$r_0$, so $U$ is monotonically increasing, i.e., $U'\geq0$. \Eref{U of p} then implies $p'\leq0$. (This is really the maximum
principle for $\Delta U=\mu+3p $.)

Now we have the following possibilities:

\begin{description}
\item \hspace{-0.2cm} (1) The pressure is positive for all values of $r$. Then the fluid fills all the space and it is not possible to extend the solution because radial ($r$-)lines have an infinite proper length (cf equations (\ref{the original metric}), (\ref{Integration of U})).

\item \hspace{-0.2cm} (2) The pressure vanishes at some value of $r$. Then we show in the
next lemma that a unique vacuum solution can be joined to the
inner solid solution.
\end{description}

\noindent {\bf Lemma 3:} {\it If $p(R)=0$, a unique vacuum solution---a particular
Levi-Civita solution---can be joined to the fluid cylinder.}

\vspace{0.3cm}

\noindent {\bf Proof:} We first show that $U(R)$ is finite. For equations of
state with a positive boundary density, $\mu_b = \mu(p=0)$,
this is obvious from equation \eref{Integration of U}. If however, the boundary density
vanishes, the integral in equation \eref{Integration of U} could diverge for $p\to 0$.
If this happens, we use the fact that $U,B,C,U',$ and $(BC)'$ are positive. Therefore, from equation \eref{Only BC and U Equation} we have
\begin{equation}
U'' <4\pi G(\mu+3p)\ ,
\end{equation}
so that
\begin{equation}
U'(r)=\int_0^r U''(r')dr<\int_0^r 4\pi G(\mu+3p)dr\ .
\end{equation}
This shows that $U'$ is bounded for $0\le r\le R$ if the EOS is monotonic (in fact, it is sufficient to assume that $\mu(p)$ is bounded on $[0,p_c]$). The same holds
for $U$. Furthermore, the monotonicity of $U$ and the upper bound on
$U''$ imply that $U,\ U'$ have limits at $R$. As $U,\ U'$ have limits at $R$, we can consider equations (\ref{equationFF}) and (\ref{equationZZ}) as linear equations for $B$ and $C$ with given coefficients. This implies that $B$, $B'$, $C$ and $C'$ have limits at $R$.

All field equations remain meaningful for $\mu=p=0$. Therefore,
the constants $B(R)$, $B'(R)$, $C(R)$, $C'(R)$, $U(R)$, $U'(R)$
provide the boundary values for the vacuum field equations; these
have a unique solution---a Levi-Civita solution. Since the metric
is $C^1$ in our radial coordinate, the junction conditions at the boundary
are satisfied by construction.

Summarizing, we have shown that a barotropic EOS determines a 1-parameter family of global spacetimes.

As mentioned in the introduction, solutions with constant density, the analogues of the interior
Schwarzschild solutions, have so far not been found as explicit
exact solutions. In section \ref{Numerics} examples of cylinders with
$\mu$=constant are constructed numerically. The following theorem
shows that a general constant-density cylinder always has a finite radius.

\vspace{0.3cm}

\noindent {\bf Theorem:} {\it \label{Theorem} For a solution with a regular axis, a barotropic EOS and a positive boundary density $\mu(p=0)=\mu_b>0$, there is a finite radius $R$ with $p(R)=0$.}

\vspace{0.3cm}

\noindent {\bf Proof:} Suppose that $p$ is positive for all $r$. \Eref{Integration of U} implies that $U$ has a limit for $r\to\infty$. We want to show
first that $U'\to 0$. If this is not the case, then there is a
sequence $r_n\to\infty$ with $U'(r_n)=a>0$. Since $U'$ is
integrable, the peaks where $U'$ reaches the value
$a$ repeatedly must become ever narrower with increasing $r$'s.
Therefore, there must be values of $r$ where there are arbitrary
high positive and negative values of $U''$. However, equation \eref{Only BC and U Equation} shows that $U''$ is bounded and we have a contradiction. Thus, $U'\to 0$.

We can rewrite equation \eref{equationTT} as
\begin{equation}
(BCU')'=4\pi G(\mu+3p)e^{-2U}BC\ ,
\end{equation}
so that by integration we get
\begin{equation}
U'={1\over BC}\int_0^r 4\pi G (\mu+3p)e^{-2U}BC dr\ .
\end{equation}
Eliminating the pressure through equation \eref{Only BC Equation}, we obtain
\begin{equation}\label{Integral for U'}
U'(r)={1\over BC}\int_0^r 4\pi G \mu e^{-2U}BC dr+{1\over
BC}\int_0^r
 {3\over8}(BC)'' dr\ ,
\end{equation}
or
\begin{equation}\fl\label{Per Parts Integral for U'}
\begin{array}{rcl}
U'(r) & = & {1\over BC} \int_0^r 4\pi G \mu e^{-2U}BC dr+{1\over BC}
 {3\over8}(BC)'(r)-{1\over BC}
{3\over8}(BC)'(0)= \\
\\
& = & {1\over BC} \int_0^r 4\pi G \mu e^{-2U}BC dr+{1\over BC}
 {3\over8}((BC)'(r)- 1/\mathcal{C}^*).
\end{array}
\end{equation}
Due to equation \eref{Constraint}, we further have
\begin{equation} \label{nitty gritty}
\fl (BC)'-\frac{1}{\mathcal{C}^*}=CB'+BC'-\frac{1}{\mathcal{C}^*}=CB'-BC'+2BC'-\frac{1}{\mathcal{C}^*}=2BC'>0.
\end{equation}
If $BC$ is bounded, the first integral gives a positive
contribution to $U'$ and we have a contradiction with $U'\to0$
because the second term in (\ref{Integral for U'}) is also positive.
For $BC$ unbounded, we calculate the limit of the integral for
$r\to\infty$ by l'Hospital rule as
\begin{equation}
\lim_{r\to\infty}{4\pi G \mu e^{-2U}BC\over (BC)'}\ .
\end{equation}
Now for $r\to\infty$ either both terms in equation \eref{Per Parts Integral for U'} (which are positive) are finite, or, if one vanishes, the other
diverges and we again have a contradiction with $U'\to 0$. Hence,
$p$ has to vanish at some finite radius. Note that equations (\ref{Per Parts Integral for U'}), (\ref{nitty gritty}) and (\ref{Constraint}) imply that this conclusion is valid for both a regular axis ($\mathcal{C}^*=1$) and for an axis with a cosmic string ($\mathcal{C}^* \not= 1$).

\section{Newtonian limit}
For a systematic treatment of the Newtonian limit, we refer to \cite{Ehlers1, Ehlers2}. Above, we have already chosen our variables in such a way that the field equations have a limit for a diverging velocity of light, i.e., for $\lambda \rightarrow 0$. For $\lambda=0$, equations (\ref{equationTT}--\ref{equationZZ}) imply
\begin{equation} \label{Newtonian Limit TT}
U''BC +B'CU'+C'BU'=4\pi G \mu BC\ ,
\end{equation}
\begin{equation} \label{Newtonian Limit RR}
B'C'=0\ ,
\end{equation}
\begin{equation}\label{Newtonian Limit FF}
C''=0\ ,
\end{equation}
\begin{equation}\label{Newtonian Limit ZZ}
B''=0\ ,
\end{equation}
\begin{equation}\label{Newtonian Limit: U of p}
U'=-{p'\over\mu}\ .
\end{equation}
Equations (\ref{Newtonian Limit RR})--(\ref{Newtonian Limit ZZ}) show that the conformal 3-space is flat in the
Newtonian limit. To adapt the coordinates to the axial symmetry
and the regularity of the axis (cf equations (\ref{Initial conditions})), we have to choose
$B=r$, $C=1$. \Eref{Newtonian Limit TT} then becomes the Poisson equation for the Newtonian gravitational potential:
\begin{equation}
U'' +{1\over r}U'=4\pi G \mu.
\end{equation}

The field equations depend on $\lambda$ in such a way that it
is easy to demonstrate the existence of families of exact
solutions that have a Newtonian limit. Suppose we choose an
EOS that is independent of $\lambda$. The theorem
in \ref{Regular Axis}, which is used to obtain solutions with a regular
axis, also shows that these solutions depend smoothly on $\lambda$
and that the limiting solution satisfies the equations for
$\lambda=0$. The same is true for the extension of the solution up
to the radius where the pressure vanishes.

Suppose we have such a $\lambda$-family of solutions with finite radii $R_\lambda$ for all $\lambda$'s. Integration of equation \eref{equationTT} up to the boundary gives
\begin{equation}\label{Integration of U'BC}
(U'BC)(R_\lambda)=G4\pi \int_0^{R_\lambda}(\mu+\lambda
3p)BCe^{-2\lambda U}dr\ .
\end{equation}
The constant
\begin{equation}\label{Definition of Tolman Mass}
m_T=2\pi \int_0^{R_\lambda}(\mu+\lambda 3p)BCe^{-2\lambda U}dr
\end{equation}
is the Tolman mass (see, e.g., \cite{Bonnor, Philbin}). Since all metric
coefficients have limits for $\lambda\to0$, equation \eref{Definition of Tolman Mass} implies that
the limit of $m_T$ is the Newtonian mass per unit length
\begin{equation}
\lim_{\lambda\to0}{m_T}=2\pi\int_0^{R_0}\mu r dr\ .
\end{equation}
The corresponding results for Newtonian cylinders and spheres are summarized in \ref{Newtonian Full Cylinders and Shells}.
\section{Joining the inner fluid solution and the outer Levi-Civita solution}
Next, we want to demonstrate explicitly how a fluid solution,
with the property that the pressure vanishes at some fixed $r=R$ such that all metric coefficients
and their first derivatives are finite there, can be
uniquely matched to a vacuum solution given by functions (\ref{LC with 5 constants})\footnote{In the following, we need to distinguish between $L$ of (\ref{LC with 5 constants}), $\rho_0$ of (\ref{LC with 3 constants}), $R$ of (\ref{LC without C}) and the coordinate $R$, proper $R_p$ and circumferential $R_c$ radii of the cylinder defined by equations (\ref{Proper Radius Definition}), (\ref{Circumferential Radius Definition}). In this section again $\lambda = 1/c^2 = 1$.}.

This is obvious because all the equations are regular for
$p=0$. Hence we can just take the values of the metric and
their first radial derivatives as the initial values for the vacuum
field equations and obtain a unique solution satisfying the
matching conditions (the metric and its first derivatives continuous) by
construction. Let us note that, as we have checked, we obtain the same results by requiring that (i) the cylinder's proper circumference is the same as measured from both sides of the surface and (ii) the surface energy-momentum tensor (calculated using the general Israel formalism \cite{Israel}) vanishes.

Equations \eref{Constraint} and \eref{LC with 5 constants} imply
\begin{equation}
\delta = \frac {L} {1-2\gamma} \frac{1}{\mathcal{C}^*}.
\end{equation}
On the surface, we have to satisfy equations (\ref{LC with 5 constants}): on the left-hand sides we have the metric potentials obtained by
integration within the cylinder, from these we now determine the five constants on the right-hand side.

Denoting $F(r) \equiv B(r)C(r)$, we have on the surface $F\:'(R)= \delta/L=1 / \mathcal{C}^*(1-2\gamma)$ and thus $\gamma = (\mathcal{C}^* F\:'-1)/(2\mathcal{C}^*F\:') = m^2/(1+m^2)$. Using this, we readily find
\begin{equation}
m = \sqrt {\frac {\mathcal{C}^*F\:'-1} {\mathcal{C}^*F\:'+1}},
\end{equation}
where $F'$ is evaluated at $R$. It is elementary to show that the outer Levi-Civita parameter $m$ is bounded by $m \in [0,1)$. Using the expression for the Tolman mass $m_T = 2 \pi \int_0^R e^{-2U} BC (\mu+3p) dr$, we find
\begin{equation}
\fl \hspace{1cm} 2m_T=U'BC=\frac{\delta}{L}\sqrt{\gamma(1-\gamma)} = \frac{1}{1-2\gamma}\frac{1}{\mathcal{C}^*}\frac{m}{1+m^2} = \frac{1}{\mathcal{C}^*}\frac{m}{1-m^2}.
\end{equation}
Thus, we obtain an explicit relation between the Levi-Civita parameter, $m$, of the outer metric, the axis conicity, $\mathcal{C}^*$, and the Tolman mass, $m_T$, of the fluid cylinder:
\begin{equation}
m_T = \frac {1} {2\mathcal{C}^*} \frac {m} {1-m^2}.
\end{equation}
Using the external metric in the form (\ref{LC without rho_0}), we find the conicity parameter from the continuity of the metric and its radial derivative. Simple calculations yield
\begin{equation}\label{Conicity Equation}
\mathcal{C} = \frac {1} {B'} \left( \frac {B} {B' e^U} \right)^\frac {\mathcal{C}^* F' -1} {\sqrt {(\mathcal{C}^* F')^2-1} -2\mathcal{C}^* F'},
\end{equation}
where $B, B', U$ and $F'$ are evaluated at $R$, the axis conicity $\mathcal{C}^*$ is constant. Consequently, the conicity parameter, $\mathcal{C}$, of the outer metric is determined by the values of the metric potentials on the surface independently of the EOS. In the case of incompressible fluid, the coefficient $\mbox{e}^U$ on the surface is given explicitly by (\ref{Integrated U}). Since the surface values of the metric, for a given EOS, are determined by the central pressure $p_c$, both $m$ and $\mathcal{C}$ are, in fact, given by $p_c$ and by $\mathcal{C}^*$ in case of a general cylinder. Even if the axis is regular ($\mathcal{C}^*=1$), the external conicity parameter $\mathcal{C} \not= 1$ in general.

\section{Cylinders of incompressible fluid: analytic approach and numerical results}\label{Numerics}
The set of the ordinary differential (field) equations (\ref{equationTT})--(\ref{equationZZ}) can be simplified by converting it into the dimensionless form:
\begin{eqnarray}
\label{Dimensionless Equations-1}{d \over ds} &N = & {Q \over H},\\
\label{Dimensionless Equations-2}{d \over ds} &Q = & (\chi+3\Pi) {H\over N} + {{\dot N}^2 H \over N},\\
\label{Dimensionless Equations-3}{d \over ds} &\Pi = & - (\chi+\Pi) {\dot N \over N},\\
\label{Dimensionless Equations-4}{d^2 \over ds^2} &{H} = &8 {\Pi H \over N^2},\\
\label{Dimensionless Equations-5}{d^2 \over ds^2} &{K} = &-{K
{\dot N}^2 \over N^2} + 2 {\Pi K \over N^2}.
\end{eqnarray}
Here the dimensionless radial coordinate $s=\beta r$, where $\beta^2 = 4\pi G \lambda
\mu_0$, with $\mu_0$ being a characteristic (e.g., the central) fluid density; the dimensionless fluid density is $\chi = \mu/\mu_0$ and the dimensionless pressure $\Pi = \lambda p / \mu_0$ (this is not an EOS). The dimensionless metric functions are given in terms of the original metric functions $U, B, C, F = BC$ as $N=e^{\lambda U}$, $K = \beta B$, $H = \beta F$, $Q = {\dot N} H$ and the dot $\dot{} = d/ds$. When considering an
incompressible fluid, we have $\chi = 1$, the solution of the equations then depends
only on the central pressure $\Pi_c=\lambda p_c/\mu_0$, which
enters into the solution through the initial conditions $N(0)=1$,
$Q(0)=0$, $\Pi(0)=\Pi_c$, $H(0)=0$, $\dot H(0)=1$, $K(0)=0$,
$\dot K(0)=1$. From now on, we assume $\mathcal{C}^*=1$ on the axis, i.e., we consider cylinders without a cosmic string along the symmetry axis. These initial conditions lead to a solution regular at $s=0$ that yields $Q/H \rightarrow 0$ and thus no
problems with `$0/0$' arise during the numerical integration of the differential equations (cf equation (\ref{Dimensionless Equations-1})). For a rigorous proof of the existence and uniqueness of a smooth solution in a neighborhood of the axis, see \ref{Regular Axis}.

Before turning to the numerical results on the fully relativistic cylinders of incompressible fluid, it is interesting to derive relativistic corrections to the Newtonian cylinders of incompressible fluid analytically. For small radii the analytic expressions agree well with fully relativistic numerical calculations.

In the analytic approach we make Taylor expansions in the dimensionless radial distance $s$ around the origin (the axis) $s=0$ of all functions entering the field equations (\ref{Dimensionless Equations-1})-(\ref{Dimensionless Equations-5}), in which, for incompressible fluids, we put $\chi=1$. Combining the results we can express the dimensionless pressure as a series in $s$ in which the coefficients are uniquely determined by the value of the pressure at the center:
\begin{eqnarray} \label{Taylor_of_Pressure}
\Pi(s) & = & \Pi_c - \frac {1}{4} \,(1 + 3\,\Pi_c)\,(1 + \Pi_c)\,s^2 + \nonumber\\
&& + \frac {1}{192} \,(61\,\Pi_c + 21)\,(1 + 3 \,\Pi_c)\,(1 + \Pi_c)\,s^4 + O(s^6) \; .
\end{eqnarray}
Note that this expansion in $s$ can also be understood as an expansion in the parameter $\lambda = 1/c^2$ since $s^2=(4\pi G \mu_0) \lambda r^2$. Hence, it also yields relativistic corrections to the corresponding Newtonian expression (given by the first two terms in the following series---see \ref{Newtonian Full Cylinders and Shells}):
\begin{eqnarray} \label{Newtonian Expansion}
p(r) & = & p_c - \pi G \mu_0^2 r^2 + \pi G \mu_0 r^2 ({7 \over 4} \pi G \mu_0^2 r^2 - 4 p) \lambda + \nonumber\\
& + & \pi G r^2 (-{317 \over 90} \pi^2 G^2 \mu_0^4 r^4 + {145 \over 12} \pi G \mu_0^2 r^2 p - 3 p^2) \lambda^2 + O(\lambda^3).
\end{eqnarray}
The expansion (\ref{Taylor_of_Pressure}) approximates well the numerical results for $s \ll 1$.
\begin{figure}[h]
\begin{center}
\epsfxsize=9cm 
\epsfbox{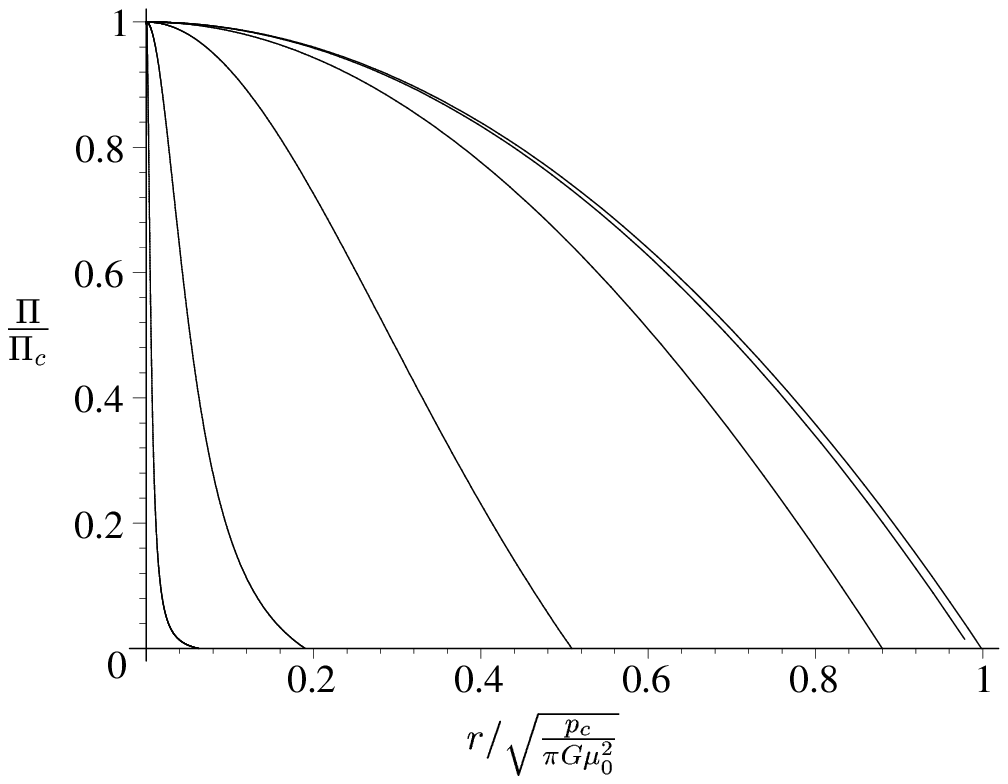}
\end{center}
\caption{\label{Pressure Taylor Expansions} Pressure fall-off as a function of the distance from the axis. The individual curves from upper right to lower left correspond to the values of the central pressure $\Pi_c = 0.001, 0.01, 0.1, 1, 10$ and $100$. The vertical axis gives the ratio of the pressure at a given point to the axis pressure $\Pi/\Pi_c$. The horizontal axis gives the ratio of the coordinate distance $r$ from the axis to the Newtonian radius of a cylinder with the same axis pressure: $r / R_{ \mbox{{\tiny Newtonian}}} = r / \sqrt {p_c / \pi G \mu_0^2}$. The central pressure increases from upper right to lower left.}
\end{figure}
Assuming that the dimensionless central pressure $\Pi_c \ll 1$, we can invert the expansion (\ref{Taylor_of_Pressure}) to find the location of the cylinder's surface, i.e., such value $s=S$ at which $\Pi(S)=0$. Notice that $\Pi_c=\lambda p_c/\mu_0 \ll 1$ is indeed valid for not very relativistic pressures.

There are two physical (geometrical) radii of a relativistic cylinder---the proper radius $R_p$ and the circumferential radius $R_c$, defined by
\begin{eqnarray}
R_p & = & \int^{R}_0 \sqrt{g_{rr}}dr = \frac {1} {\sqrt{4 \pi G\lambda\mu_0}} \int_0^S \frac {ds} {N},\label{Proper Radius Definition}\\
R_c & = & \sqrt{g_{\varphi\varphi}|_{r=R}} = \frac {1} {\sqrt{4 \pi G\lambda\mu_0}} \left. \frac {K} {N} \right|_{s=S}.\label{Circumferential Radius Definition}
\end{eqnarray}
Using the expansions of the metric functions $K, N$ in $s^2$ and substituting for $S$ from the condition $\Pi(S)=0$ determined by inverting equation \eref{Taylor_of_Pressure}, we arrive at the following results:
\begin{eqnarray} \label{Series for Radii 1}
R_p & = & \sqrt{p_c \over \pi G \mu_0^2}\left(1-{35\over 24}\Pi_c+{18347\over 5760} \Pi_c^2+ O(\Pi_c^3)\right),\\
R_c & = & \sqrt{p_c \over \pi G \mu_0^2}\left(1-{17\over
8}\Pi_c+{6415\over 1152} \Pi_c^2+ O(\Pi_c^3)\right). \label{Series for Radii 2}
\end{eqnarray}
Here the dimensionless central pressure plays the role of an expansion parameter; since it involves $\lambda$, the second and subsequent terms in (\ref{Series for Radii 1}), (\ref{Series for Radii 2}) determine the relativistic corrections. The first terms, of course, coincide: $R_p=R_c=\sqrt{p_c/\pi G\mu_0^2}$ is the radius of a Newtonian cylinder of incompressible fluid with density $\mu_0$ and central pressure $p_c$ (see \ref{Newtonian Full Cylinders and Shells}).

Next, defining mass per unit coordinate and proper lengths of the cylinder, $M_1$ and $M_p$, by
\begin{equation}\label{Unit_Coordinate_Length_Mass}
G \lambda M_1 = 2\pi G \lambda \int_0^R \mu BC e^{-3\lambda U} dr = \frac {1} {2} \int_0^S \frac {H} {N^3} ds,
\end{equation}
and
\begin{equation}\label{Unit_Proper_Length_Mass}
G \lambda M_p = 2\pi G \lambda \int_0^R \mu B e^{-2\lambda U} dr = \frac {1} {2} \int_0^S \frac {K} {N^2} ds,
\end{equation}
respectively, we find expansions
\begin{equation}
G \lambda M_1 = \Pi_c - \frac {15}{4} \Pi_c^2 +\frac{995}{72} \Pi_c^3 + O(\Pi_c^4),
\end{equation}
and
\begin{equation}
G \lambda M_p = \Pi_c - \frac {13}{4} \Pi_c^2 +\frac{731}{72} \Pi_c^3 + O(\Pi_c^4).
\end{equation}
Again, the first term yields the expected Newtonian result: $M_1 = M_p = \Pi_c / G \lambda = p_c / G \mu_0$. The Levi-Civita mass parameter turns out to be
\begin{equation} \label{Series for m}
m = \left. \sqrt {\frac {\dot{H} -1} {\dot{H} +1}}\: \right|_{s=S} \hspace{-0.5cm} = {2\Pi_c} \left( 1- \frac {7} {4} \Pi_c+{185 \over 72} \Pi_c^2+ O(\Pi_c^3)\right),
\end{equation}
so that in the lowest order $G \lambda M_1 = G \lambda M_p = m/2$, as is also the case for infinitely thin cylindrical shells \cite{BZ}.

The conicity parameter, $\mathcal{C}$, outside the cylinder is also uniquely determined by the central pressure as it follows from equation \eref{Conicity Equation} with $\mathcal{C}^*=1$:
\begin{equation} \label{Series for c}
\mathcal{C} = 1 + \Pi_c^2 \left( 2 \ln \frac {\beta^2} {4 \Pi_c} -1 \right) - \Pi_c^3 \left( 3 \ln \frac {\beta^2} {4 \Pi_c} - \frac {41} {3} \right) + O(\Pi_c^4).
\end{equation}
The conicity parameter starts to deviate from 1 only by terms of order $O(\Pi_c^2 \ln \Pi_c)$.

Let us now turn to the numerical results for fully relativistic cylinders of incompressible fluid. The numerical integration of the system (\ref{Dimensionless Equations-1})-(\ref{Dimensionless Equations-5}) (with $\chi=1$) enables one to find first the radial distribution of the pressure and then the proper and circumferential radii of the cylinders, as well as other physical quantities, all being determined uniquely by the central dimensionless pressure $\Pi_c=p_c/\mu_0c^2$.

The dependence of the pressure on the dimensionless distance $s$ from the axis is illustrated in figure \ref{Pressure Taylor Expansions}. The higher the central pressure, the more the resulting curve deviates from the Newtonian case. The Newtonian curve is represented by a parabola $y=1-x^2$ for any central pressure. For low central pressures, the curve is very close to this parabola; as the central pressure increases, it drops faster than in the Newtonian case and the surface is thus closer to the axis.
\begin{figure}[ht]
\begin{center}
\epsfxsize=9cm 
\epsfbox{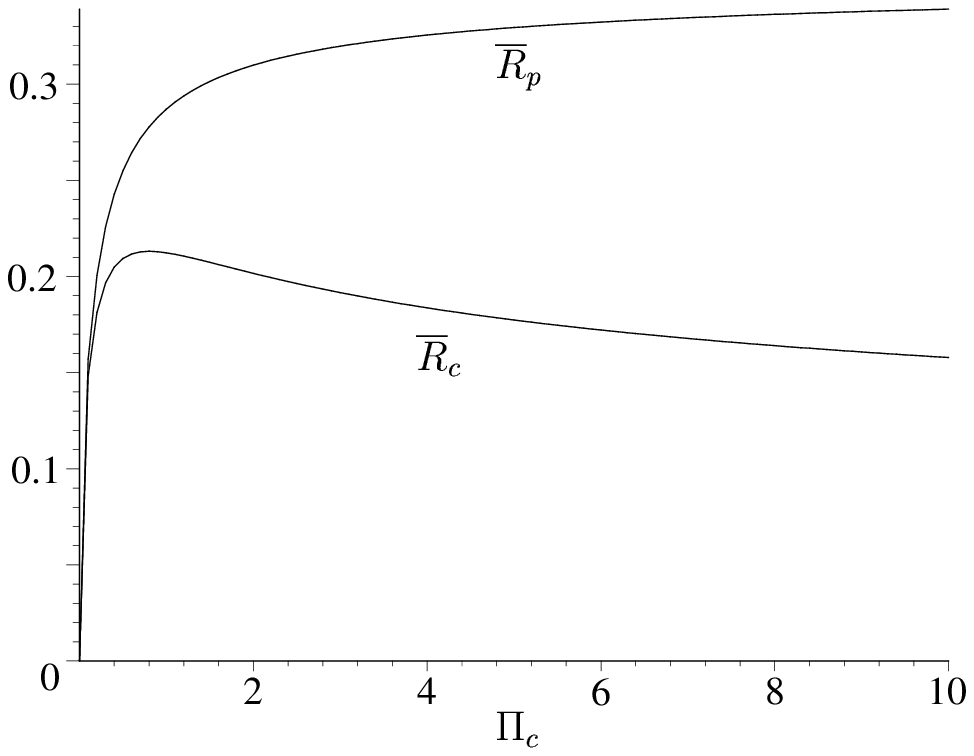} 
\end{center}
\caption{\label{Full Cylinders-radii} The dependence of the dimensionless proper radius of the cylinder $\overline{R}_p=\sqrt{G\lambda\mu_0} R_p$ and the dimensionless circumferential radius $\overline{R}_c=\sqrt{G\lambda\mu_0} R_c$ of an incompressible fluid cylinder on the dimensionless central pressure $\Pi_c = \lambda p_c/\mu_0$. The circumferential radius $\overline{R}_c$ has only a bounded range of values.}
\end{figure}

Perhaps the most interesting result can be seen in \fref{Full Cylinders-radii}. Here the dimensionless proper and circumferential radii, $\overline{R}_p$ and $\overline{R}_c$, of the cylinders are illustrated for cylinders parameterized by the central pressure $\Pi_c$. Curiously enough, while the proper radius increases with increasing $\Pi_c$, the circumferential radius starts to decrease for higher central pressures, though with still `physical' values of $\Pi_c<1$; the maximum of the curve $\overline{R}_c(\Pi_c)$ gives $\overline{R}_c \approx 0.213243$ and occurs at $\Pi_c \approx 0.8$ (see \fref{Full Cylinders-radii}). For any finite value of $\mu_0=$ constant $>0$ and any finite $\Pi_c>0$, the coordinate radius of the cylinder, $R$---and, correspondingly, also $R_p$ and $R_c$---is finite in accordance with the theorem in section \ref{Global Solutions}.

\begin{figure}[ht]
\begin{center}
\epsfxsize=9cm 
\epsfbox{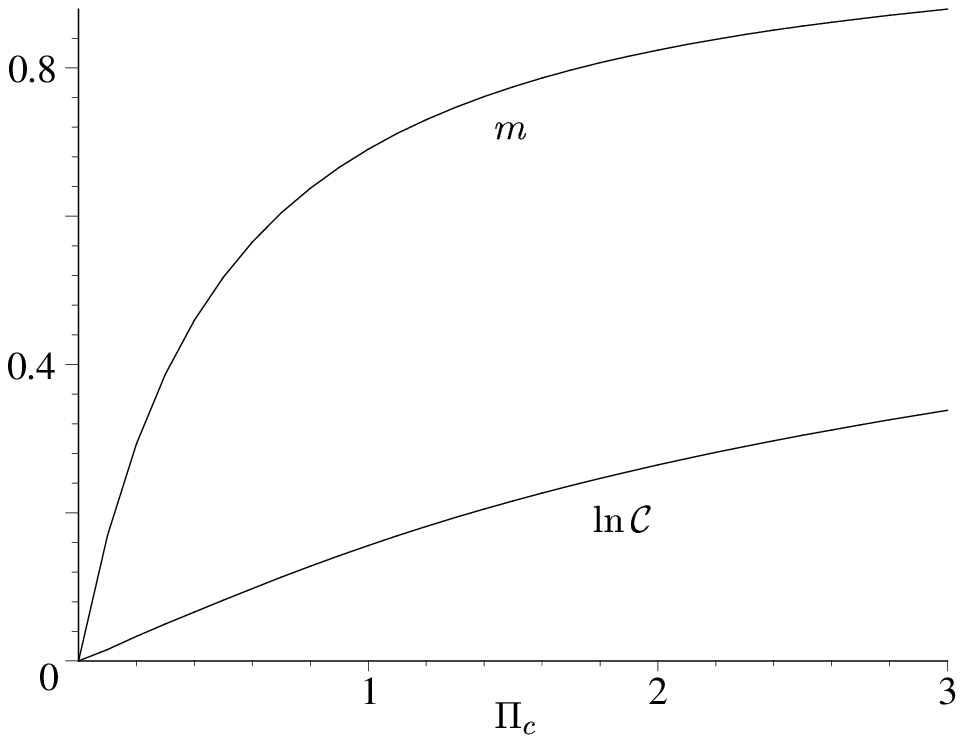} 
\end{center}
\caption{\label{Full Cylinders-m and C} The external Levi-Civita parameter, $m$, and the external conicity parameter, $\ln(\mathcal{C})$, as functions of the dimensionless central pressure, $\Pi_c = \lambda p_c/\mu_0$, for incompressible fluid ($m$ can attain only values within $[0,1)$). We set $\beta^2 = 4 \pi G \lambda \mu_0 = 1$.}
\end{figure}

As discussed below (\ref{Conicity Equation}), knowing the cylinder radius, we can use the matching conditions to calculate the Levi-Civita mass parameter, $m$, and the conicity parameter, $\mathcal{C}$, of the vacuum spacetime outside the cylinder for any given $\Pi_c$. The resulting curves are illustrated in \fref{Full Cylinders-m and C}. The Levi-Civita mass parameter, characterizing the curvature of the vacuum spacetimes (see equation \eref{LC with 3 constants}), increases from its flat-space value, reaching the magnitude of $m \approx .69$ for $p_c= \mu_0 c^2$, and approaching $m = 1$ for the extreme central pressures $\Pi_c \gg 1$. Interestingly, the cylinders with still relatively low central pressures are so relativistic that they produce Levi-Civita solutions with $m>1/2$, in which there are no circular timelike geodesics (cf also \cite{Bonnor-Interpretation} below Equation 14). For an analogous phenomenon in case of static cylindrical shells and their Levi-Civita fields, see \cite{BZ}.

In the physical region of the pressures, $p_c \leq \mu_0 c^2$, one can find, by numerical interpolation, the following nice analytical approximations (all with a relative accuracy better than $10^{-3}$ up to $\Pi_c = 1$) to the numerical curves illustrated in figures \ref{Full Cylinders-radii}, \ref{Full Cylinders-m and C}:
\begin{eqnarray}
R_p & \sim & \sqrt{p_c \over \pi G \mu_0^2} \left( 1-\Pi_c{532 + 36\Pi_c\over 368+788\Pi_c}\right),\\
R_c & \sim & \sqrt{p_c \over \pi G \mu_0^2} \left( 1-\Pi_c{473 + 21\Pi_c\over 225+567\Pi_c}\right),\\
m & \sim & {2\Pi_c} \left( 1-\Pi_c{733 - 24\Pi_c\over 413+670\Pi_c} \right),\\
\mathcal{C} & \sim & \left( 1 + \Pi_c^2 \frac {41-3 \Pi_c} {5+220 \Pi_c} \right) \beta^{ \Pi_c^2 \frac {168+15 \Pi_c + 5 \Pi_c^2} {42+68\Pi_c + 184 \Pi_c^2} }.
\end{eqnarray}

\begin{figure}[ht]
\begin{center}
\epsfxsize=9cm 
\epsfbox{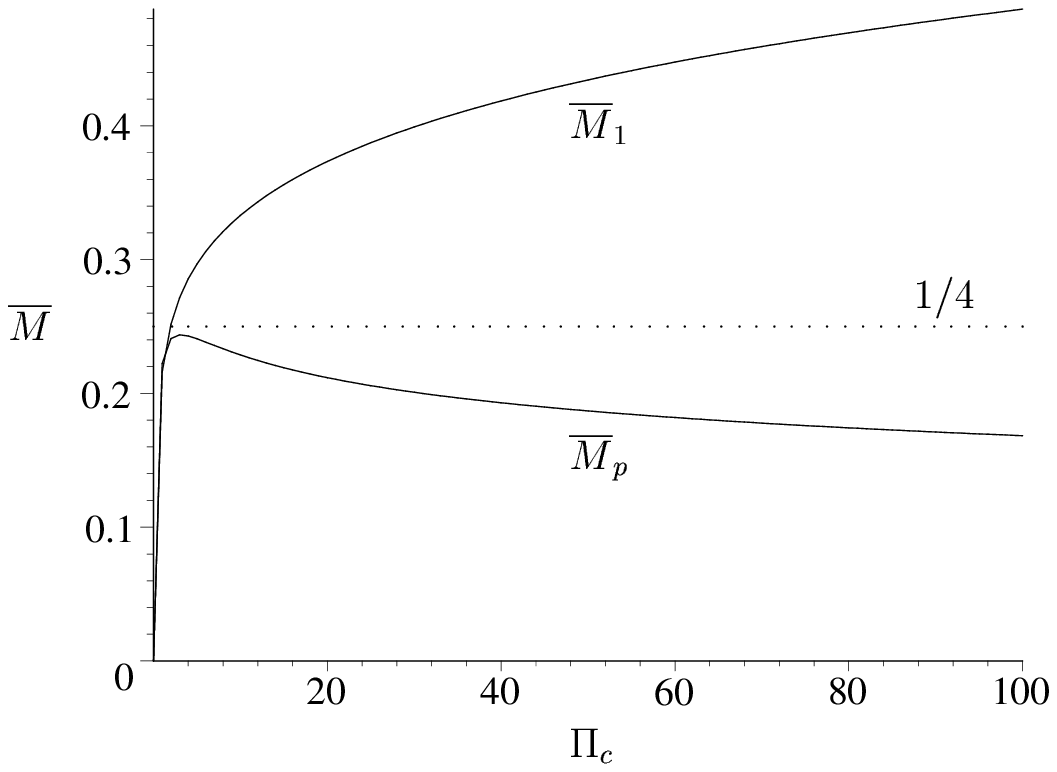} 
\caption{\label{Unit Length Masses} Dimensionless masses per unit coordinate and unit proper lengths of the cylinders, $\overline{M}_1 = G \lambda M_1$ and $\overline{M}_p = G \lambda M_p$, as functions of the dimensionless central pressure $\Pi_c$. The dimensionless mass per unit proper length $\overline{M}_p$ has an upper bound of $1/4$.}
\end{center}
\end{figure}

\begin{figure}[ht]
\begin{center}
\epsfxsize=9cm 
\epsfbox{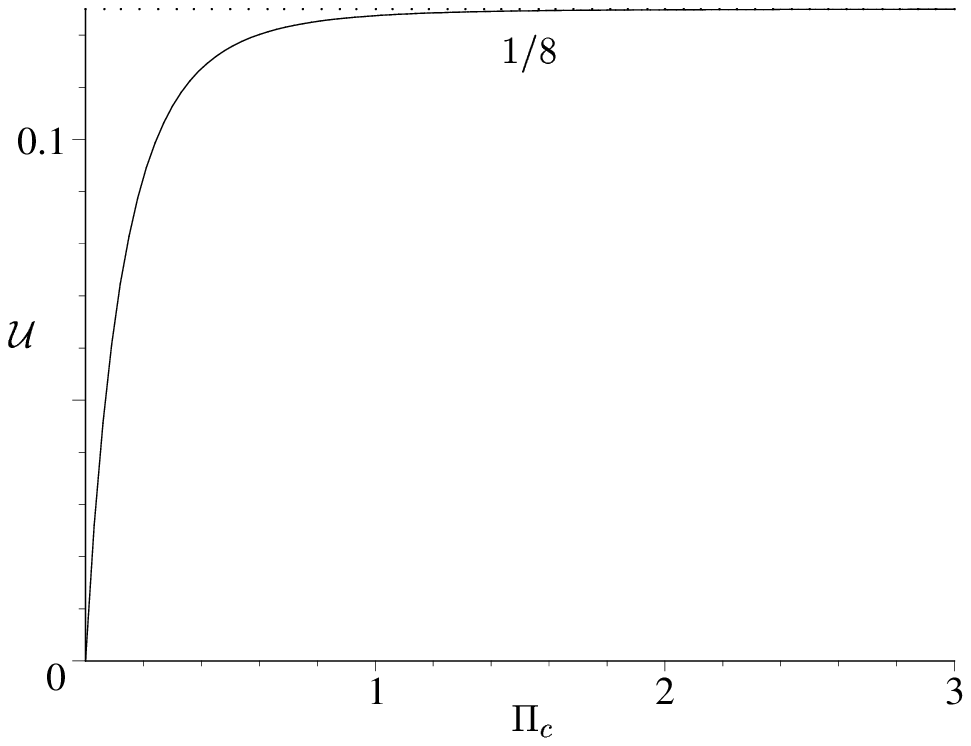} 
\end{center}
\caption{\label{C Energy} C-energy scalar $\mathcal{U}$ evaluated on the surface of the cylinders as a function of the dimensionless central pressure $\Pi_c$. As can be seen, this quantity is bounded from above by $1/8$.}
\end{figure}

It is not obvious what expression to use as the unit-length mass of the cylinders. There are several choices. One can use the Vishveshwara-Winicour definition \cite{VW} employing Killing vector fields outside the cylinders (in the outer Levi-Civita spacetime) and yielding
\begin{equation}
M_{VW}= \frac {m} {2}.
\end{equation}
We find $M_{VW} \in [0,1/2)$. This value crosses the $1/4$ limit \cite{BZ} already for the central pressure $\Pi_c$ well below $1$. We can also use the Tolman mass
\begin{equation}
\fl \hspace{1cm} m_T = 2\pi \int_0^R(\mu+\lambda 3p)BCe^{-2\lambda U}dr = \frac {1} {2G} \left. U'BC \right|_{r=R} = \frac {1} {2 G \lambda } \left. \frac {Q} {N} \right|_{s=S},
\end{equation}
which is not bounded from above---solutions with unbounded $m_T$ have also been found analytically \cite{Philbin}. We can use mass per unit coordinate length, $M_1$ \eref{Unit_Coordinate_Length_Mass}, with no upper bound (see \fref{Unit Length Masses}).
There are two more expressions that do exhibit a limited interval of values: mass per unit proper length of the cylinder, $M_p$ \eref{Unit_Proper_Length_Mass}, with $G \lambda M_p \leq 1/4$ as shown in \cite{Anderson} (see \fref{Unit Length Masses}), and Thorne's C-energy scalar defined by using the symmetries of the spacetime \cite{Thorne} as
\begin{equation}
\mathcal{U} = \frac {1} {8} \left[ 1 - \frac {A_{,\mu} A^{,\mu}} {4 \pi^2 | \frac {\partial} {\partial z} |^2} \right] = \frac {1} {8} \left[ 1-\left( \frac {KN^2} {H} \frac {d}{ds} \left[ \frac {H} {N^2} \right] \right)^2 \right],
\end{equation}
where $A = 2 \pi BC e^{-2U}$ is the area of the cylindrical belt given by $r=$ constant, $z \in [0,1]$, $\varphi \in [0,2 \pi)$. We find $\mathcal{U} \leq 1/8$---in accordance with \cite{Thorne} (see \fref{C Energy}).

There is a fundamental difference between the two bounded expressions---there is no increase in $M_p$ outside the solid cylinders but there is non-zero C-energy scalar associated also with the outer vacuum Levi-Civita spacetime. The total C-energy contained within a cylinder of radius greater than the radius of the solid cylinder is given simply by the corresponding expression for the pure Levi-Civita spacetime (no integration) $\mathcal{U} = (1/8)(1 - (1-m)^4 / r^{2m^2} \mathcal{C}^2) \leq 1/8$.

\begin{figure}[ht]
\begin{center}
\epsfxsize=9cm 
\epsfbox{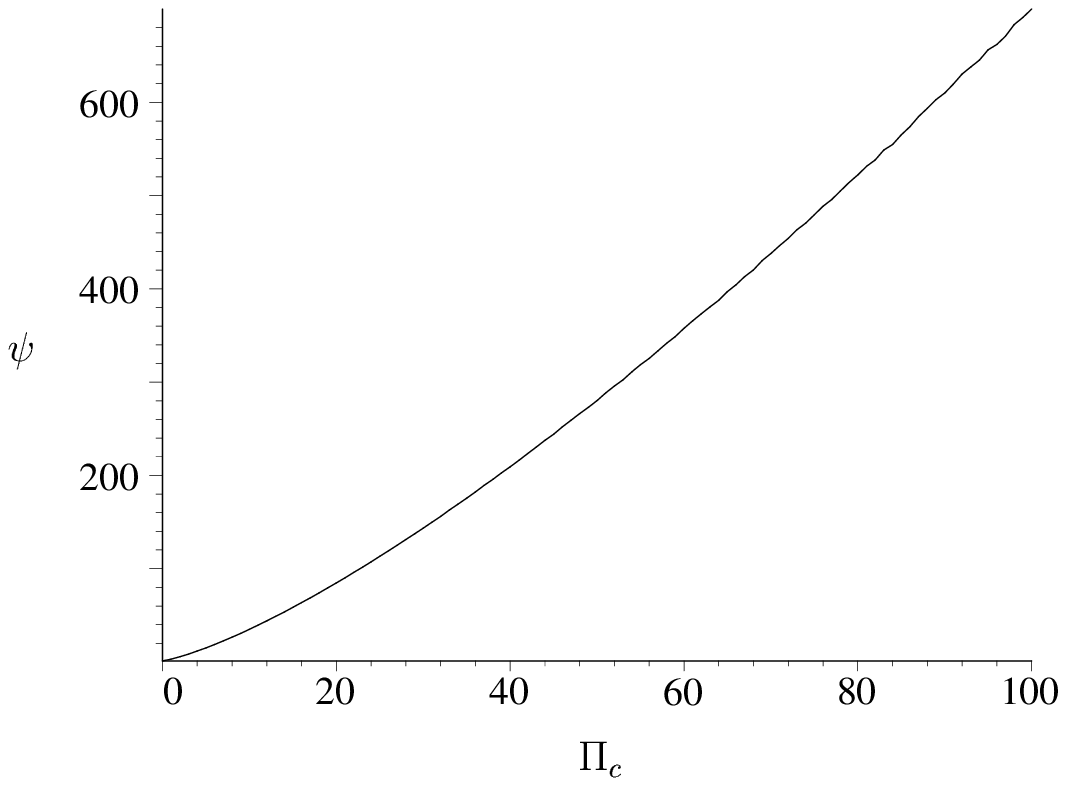} 
\end{center}
\caption{\label{Psi} The conicity characterizing quantity $\psi$ (see equation \eref{Conicity definition 2}) evaluated at the surface of the cylinder with the central pressure $\Pi_c$.}
\end{figure}

We can construct the following invariant expression characterizing the conicity of the metric
\begin{equation}\label{Conicity definition 2}
\psi \equiv \lim_{r_2 \rightarrow r_1} \frac {R_p(r_2) - R_p(r_1)} {R_c(r_2) - R_c(r_1)} = 2 \frac {\sqrt {g_{ \varphi \varphi} (r_1) g_{rr} (r_1)}} {\frac {d} {dr} g_{ \varphi \varphi} (r_1)},
\end{equation}
with $R_c$ being the circumferential and $R_p$ the proper radius. In Minkowski spacetime we find $\psi=1$. For a cosmic string $ds^2= - dt^2 + dr^2 + dz^2 + \frac {1}{{\mathcal{C}}^2} r^2 d\varphi^2$, and we have $\psi = \mathcal{C}$; in a Levi-Civita spacetime we calculate $\psi = \mathcal{C} r^{m^2}/(1-m)$ (and thus $\mathcal{U} = (1/8) (1-[(1-m)/\psi]^2)$). For full cylinders, we obtain
\begin{equation}
\psi = \frac {1} {B'-BU'} = \frac {N} {N \dot{K} -K \dot{N}}
\end{equation}
---see \fref{Psi}.

\begin{figure}[ht]
\begin{center}
\epsfxsize=5cm 
\epsfbox{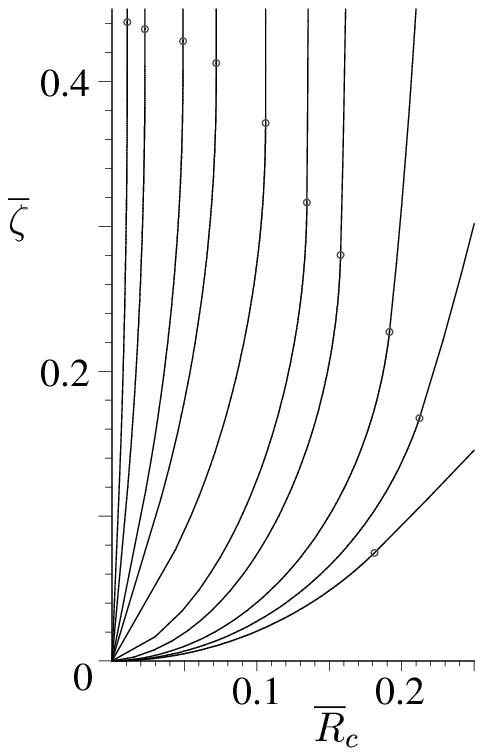} 
\end{center}
\caption{\label{Embedding} Embedding of the $z,t=$ constant surfaces according to formula (\ref{Embedding Surface}) for central pressures $\Pi_c=0.2, 1, 3, 10, 25, 100$, $1000$, $10^4$, $10^6$ and $10^8$, bottom to top. The dots on the graph indicate the position of the cylinder's surface. $\overline{R}_c = \sqrt{ G \lambda \mu_0} R_c$ is the dimensionless circumferential radius, $\overline{\zeta} = \sqrt{ G \lambda \mu_0} \zeta$.}
\end{figure}

If we embed a 2-dimensional hyperplane $t, z =$ constant with metric
\begin{equation}
ds^2 = \mbox{e}^{-2 \lambda U} \left( dr^2 + B^2 d\varphi^2 \right)
\end{equation}
into a flat Euclidian space with metric
\begin{equation}
ds^2 = d\mathcal{R}^2+ \mathcal{R}^2 d\varphi^2 + d\zeta^2,
\end{equation}
we get
\begin{equation}
\mathcal{R} = B \mbox{e}^{- \lambda U} = \frac {K} {\beta N}.
\end{equation}
\Eref{Circumferential Radius Definition} implies $\mathcal{R}= R_c$ and finally (using \eref{Proper Radius Definition})
\begin{equation}\label{Embedding Surface}
\frac {d\zeta} {dR_c} = \sqrt {\frac {1} {\left( N \frac {d} {ds} (K/N) \right)^2} -1} = \sqrt {{\frac {dR_p} {dR_c}}^2-1} \;.
\end{equation}
We conclude
\begin{equation}
dR_p^2 = dR_c^2 +d\zeta^2.
\end{equation}
Thus $R_p$ measures the length of the embedding curve, see \fref{Embedding}. It can be shown that the embedding surface never degenerates into a cylinder ($d\zeta / dR_c = \infty$) at a finite distance from the axis.
For a cosmic string, this simplifies to a cone $d\zeta / dR_c = \sqrt {\mathcal{C}^2-1}$, as expected. Further we find
\begin{equation}
\psi = \sqrt {1+ \left( \frac {d\zeta} {dR_c} \right)^2} \hspace{1cm} \mbox{and} \hspace{1cm} \frac {d\zeta} {dR_c} = \sqrt {\psi^2 -1 } \;.
\end{equation}

\begin{figure}[h]
\begin{center}
\epsfxsize=9cm
\epsfbox{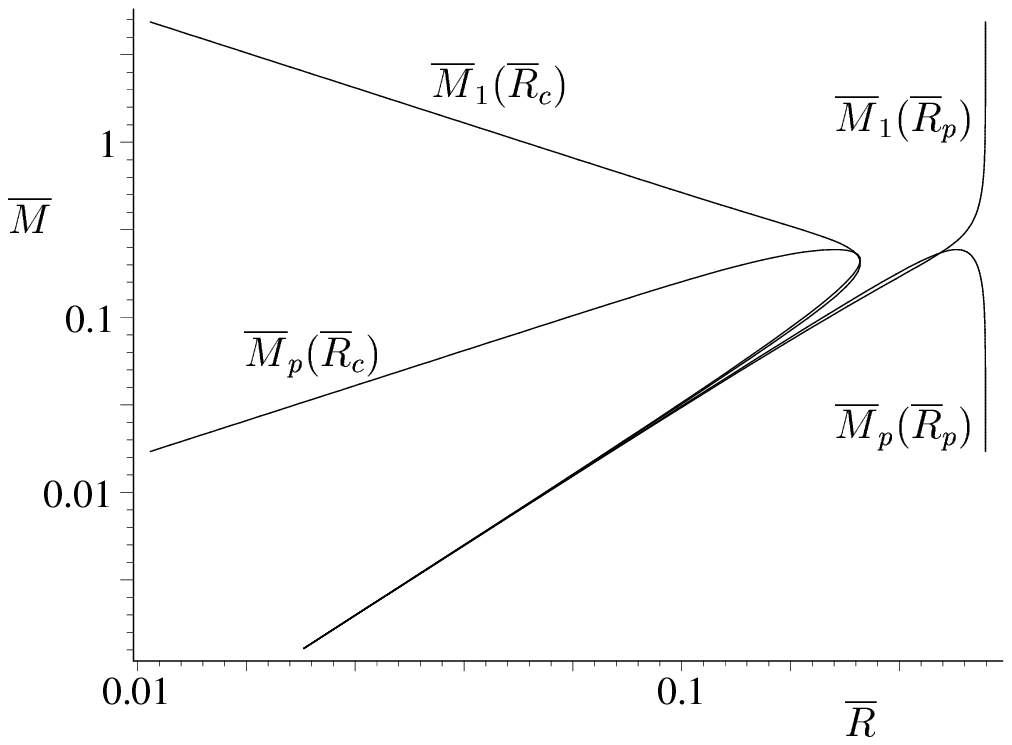}
\end{center}
\caption{\label{Mass per Unit Length} Dimensionless masses of the cylinders per unit coordinate and unit proper lengths, $\overline{M}_1 = G \lambda M_1$ and $\overline{M}_p = G \lambda M_p$, as functions of their dimensionless proper and circumferential radii, $\overline{R}_p = \sqrt{G \lambda \mu_0} R_p$ and $\overline{R}_c = \sqrt{G \lambda \mu_0} R_c$. The graph uses logarithmic scales to reveal the asymptotic behavior of the masses.}
\end{figure}

\begin{figure}[ht]
\begin{center}
\epsfxsize=9cm \epsfbox{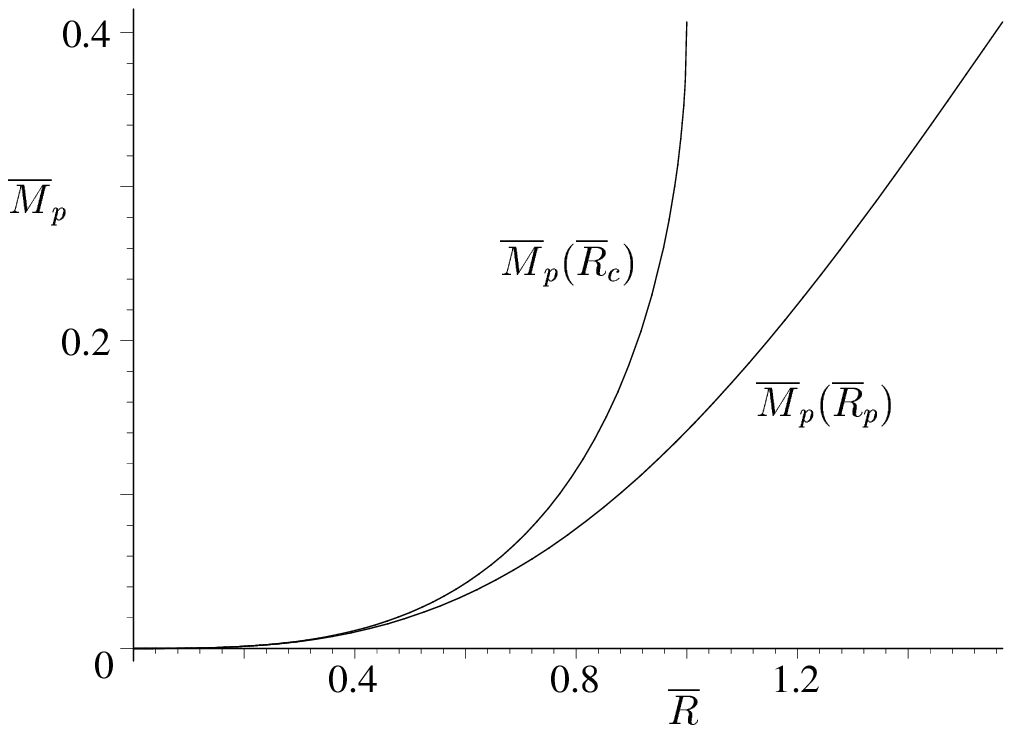}
\end{center}
\caption{\label{Full_Spheres} Dimensionless total proper mass $\overline{M}_p = \sqrt {G \lambda \mu_0} \; G \lambda M_p$ of a static spherical star of constant density $\mu_0$ as a function of its dimensionless proper radius $\overline{R}_p = \sqrt{8 \pi G \lambda \mu_0 / 3} \: R_p$ and dimensionless circumferential radius $\overline{R}_c = \sqrt{8 \pi G \lambda \mu_0 / 3} \: R_c$. In the spherical case, the circumferential radius $R_c$ coincides with the coordinate radius $R$ in the standard spherical coordinates.}
\end{figure}

Another interesting graph is a plot of unit-length mass within the cylinder as a function of the cylinder radius. We have four options: mass per unit coordinate or proper length and proper and circumferential radii. In \fref{Mass per Unit Length} we present all four quantities. The graph of the unit proper length mass as a function of the circumferential radius resembles the plot of equilibrium spherical configurations where we plot the Schwarzschild mass $M$ of the system as a function of its coordinate radius $R$. On the other hand, if we integrate the structure equations in case of a spherically symmetric static star of constant density $\mu_0$, we find
\begin{eqnarray}\label{Spherical_Star}
G \lambda M_p & = & \frac {3} {4} \left(R_p - \frac {1} {\mathcal{A}} \sin \mathcal{A} R_p \; \cos \mathcal{A} R_p \right) = \nonumber \\
& = & \frac {3} {4} \left( \frac {1} {\mathcal{A}} \arcsin (\mathcal{A}  R_c) - R_c \; \sqrt {1- \mathcal{A}^2 R_c^2} \right),
\end{eqnarray}
where $\mathcal{A} = \sqrt {8 \pi G \lambda \mu_0 / 3}$, $M_p$ is the total proper mass of the star and $R_p$, $R_c$ are its proper and circumferential radii, respectively (see \fref{Full_Spheres}). This function is similar to the first graph in \fref{Mass per Unit Length}---to the mass per unit coordinate length $M_1$ as a function of the dimensionless proper radius $\overline{R}_p$ of the cylinder.
There is, however, a fundamental difference between the spherical and cylindrical configurations---in the spherical case in standard Schwarzschild spherical coordinates, the radial component of the metric tensor depends on the density but not on the pressure. Therefore, we do not need to find the pressure as a function of the distance from the center to evaluate $M_p$. For spheres of incompressible fluid, the dependence of the proper mass on the distance from the center inside the sphere is common for all spheres and it is the same as the dependence of the total proper mass on the radius of the spheres. This is not the case for cylinders of incompressible fluid where each value of the central pressure determines a unique curve $p(r)$.
\section{Concluding remarks}
Although the main results were announced in the introductory section already, let us briefly summarize them here, emphasizing those aspects that appear to be new in the subject of cylindrical symmetry in general relativity. Our primary goal, in contrast to majority of other works, has been (i) to understand the global character of spacetimes with static perfect fluid cylinders, (ii) to study weakly gravitating cylinders and their Newtonian limit and (iii) to analyze in detail cylinders of incompressible fluid. Except for the last item, we did not start from an {\it a priori} equation of state, so our results are of a general character.

We have shown that for any smooth monotonic equation of state and for any density at the axis of symmetry there exists a unique solution in some neighborhood of the axis which is regular at the axis and can be uniquely extended to a global solution. In other words, the equation of state determines a one-parameter family of global spacetimes. If the pressure vanishes at a finite value of the radial coordinate then a unique Levi-Civita vacuum solution can be joined to the inner perfect-fluid solution. In particular, we prove that the cylinder has a finite radius if the monotonic equation of state admits a non-vanishing density for zero pressure (as is the case for incompressible fluid, for example). In general, the outside Levi-Civita solution is determined by both the mass (curvature) parameter and the conicity parameter. The need for a nontrivial conicity parameter in the external vacuum region and the fact that both the mass and conicity parameters are determined uniquely by the value of the density/pressure at the axis of the fluid cylinder do not appear to have been elucidated before.

In the Newtonian limit, we prove that the spatial metric inside the fluid cylinders is conformally flat. The relativistic Tolman mass becomes the Newtonian mass per unit length. For comparison, we also discussed cylinders and spheres of perfect fluid constructed {\it ab initio} in the Newtonian theory.

In the case of relativistic cylinders of incompressible fluid, no analytical solution is available. However, we succeeded in deriving analytic results for the relativistic corrections to the Newtonian cylinders. For weekly relativistic cylinders, the Levi-Civita mass parameter and the conicity parameter, for example, are given by equations (\ref{Series for m}) and (\ref{Series for c}). For fully relativistic cylinders, various physical quantities of interest were found by numerical integration of the field equations and they are exhibited graphically (figures 1--8 in section 6). The resulting configurations are determined uniquely by the dimensionless pressure $\Pi_c = p_c / \mu_0 c^2$ at the axis. As the central pressure increases, the pressure away from the axis decreases more rapidly so that relativistic cylinders become more compact than the Newtonian ones. A remarkable phenomenon, unnoticed so far, arises in the relativistic regime: while the (dimensionless) proper radius of the cylinders increases with increasing central pressure, the (dimensionless) circumferential radius starts to decrease for still `physical' values of $\Pi_c < 1$; the circumferential radius has only a bounded range of values. This phenomenon is nicely illustrated by the embedding diagrams of the surfaces $z, t=$constant (figure \ref{Embedding}). The external Levi-Civita mass parameter $m$ increases with $\Pi_c$, approaching $m=1$ as $\Pi_c \gg 1$, whereas the quantity $\psi$ characterising the conicity increases without limit as $\Pi_c \rightarrow \infty$. The dimensionless mass per unit proper length inside the cylinders starts to decrease for highly relativistic cylinders (it has an upper bound of $1/4$), but the analogous mass per unit coordinate length increases without bound. Thorne's C-energy increases with $\Pi_c$ and approaches $1/8$ as $\Pi_c$ becomes large.

At the end of the preceding section, a comparison of cylindrical and spherical configurations of incompressible fluid shows similarities and fundamental differences between the two cases (figures 8 and 9).

The `external' Levi-Civita conicity parameter, the importance of which has been emphasized throughout the text, takes various values depending on the `inner' cylinder, which is regular everywhere. However, to include more general situations, we also considered an infinitely thin cosmic string (see equation (\ref{Initial conditions})), i.e., a conical singularity along the axis of symmetry inside the cylinder. This produces the `axis conicity' which enters various formulae, like the Tolman mass, but it does not, for example, influence the conclusions of the theorem stating that a cylinder has a finite radius if the equation of state admits positive density at vanishing pressure.

Finally, to `end with the beginning', let us remark again that only full understanding of the static situation will enable a sufficiently thorough treatment of problems such as the interaction of cylindrical gravitational waves with static cylindrical matter configurations.
\ack
J.B. thanks the Albert-Einstein-Institute, Golm, for hospitality. J.B., T.L. and M.\v{Z}. acknowledge support by Grant No.~GA\v{C}R 202/02/0735 and Research Project No.~MSM113200004. We thank the referees for some particularly useful suggestions.
\appendix \section[]{Regularity of the axis} \label{Regular Axis}

In order to prove the existence of solutions regular on the axis we begin with the following two equations ($2 \: \times$ (\ref{equationRR}) $+ \: B \: \times$ (\ref{equationFF}) $+ \: C \: \times$ (\ref{equationZZ})) and (\ref{equationTT}) written for $F=BC$:
\begin{equation}
{F''}=8\pi\lambda^2 G e^{-2U\lambda} 4p F = V(U,\lambda)F\, ,
\end{equation}
\begin{equation}
U'' +{F'\over F}U'=4\pi G
(\mu+\lambda^23p)e^{-2U\lambda}=W(U,\lambda)\, .
\end{equation}
For a given EOS, $\mu(p)$, and the value of the pressure on the axis, $p_c>0$, the function $p(U)$ is uniquely
determined from equation \eref{U of p} since $p+\mu$ does not change the sign. The boundary values at $r=0$ are
$F(0)=U(0)=U'(0)=0, F'(0)=1$.

We want to write the above system in the form
\begin{equation}\label{Rendall Schmidt}
x{df\over dx}+ Y f =xG(x,f(x)) + g(x)\, ,
\end{equation}
where $Y$ is a constant $n\times n$ matrix and $f(x)$ a vector. For this form, the existence and uniqueness of a smooth solution in a neighborhood of $x=0$ was shown in theorem 1 in \cite{Rendall-Schmidt}, provided the matrix $Y$ has positive eigenvalues.

First we define
\begin{equation}
F(r)=r\hat f(r^2)\, ,
\end{equation}
where $\hat f(0)=1$ and $\hat f'(0)$ is finite, and obtain the
equation
\begin{equation}
U'' + {1\over r} U' + 2r {\hat f'\over \hat f}=W\, .
\end{equation}
With $x=r^2$ as radial coordinate and $\hat U(r^2)=U(r)$, we get
\begin{equation}
4x\hat U'' +4\hat U' + 4x {\hat f'\over \hat f}\hat U' =W(\hat
U,\lambda)\, ,
\end{equation}
where ${}'$ is the derivative with respect to the argument of the
function, i.e, $x$. We write this equation as a first order system
\begin{equation}
\hat U'=\hat v\, ,
\end{equation}
\begin{equation}
4x\hat v' +4\hat v + 4x {\hat f'\over \hat f}\hat v =W(\hat
U,\lambda)\, .
\end{equation}
With $\hat U(x)=x\hat u(x)$, we obtain
\begin{equation}
x\hat u' + \hat u -\hat v=0\, .
\end{equation}
Similarly, we rewrite the equations for $\hat f$ as a singular
first-order system. Defining
\begin{equation}
\hat f'=\hat g\, ,
\end{equation}
we get
\begin{equation}
4x \hat g' +6 \hat g = V\hat f\, .
\end{equation}
To make the equation singular, we define
\begin{equation}
\hat f=x\hat h+1
\end{equation}
and obtain
\begin{equation}
x\hat h'+\hat h-\hat g=0\, .
\end{equation}
Now we finally have the system in the desired form of equation \eref{Rendall Schmidt}:
\begin{equation}
x\hat v' +\hat v=- x {x\hat h'+\hat h\over x\hat h+1}\hat v
+{1\over 4}W(x\hat u,\lambda) \, ,
\end{equation}
\begin{equation}
x\hat u' + \hat u -\hat v=0\, ,
\end{equation}
\begin{equation}
x \hat g' +{3\over 2} \hat g ={1\over 4}(1+x\hat h) V(x\hat
u,\lambda)\, ,
\end{equation}
\begin{equation}
x\hat h'+\hat h-\hat g=0\, .
\end{equation}
The matrix $Y$ is
\begin{equation}
Y=
\left(\matrix{ 1 &-1& 0&0&\cr
    0 & 1& 0&0&\cr
    0 & 0& {3\over 2}& 0& \cr
    0 & 0& -1& 1& }\right)\, ,
\end{equation}
which has positive eigenvalues $1$, ${3\over 2}$.
\section[]{Static Newtonian configurations of matter} \label{Newtonian Full Cylinders and Shells}

For a Newtonian cylinder, the balance equation reads
\begin{equation}
p'+g\mu=0,
\end{equation}
where $g$ is gravitational acceleration, $p$ is pressure and $\mu$ is the fluid density (variable in general). Using the Gauss theorem, we have
\begin{equation}
p'+\frac {4\pi G \mu} {r} \int_0^r \mu r dr=0.
\end{equation}
Since there appears an integral quantity, namely the total mass contained within the cylinder up to radius $r$, we obtain a second-order differential equation
\begin{equation}\label{Balance Equation}
4 \pi G r \mu^3 = rp'\mu'-\mu p' -\mu r p''.
\end{equation}
For incompressible fluid of density $\mu_0$, pressure $p$ is a quadratic function of distance from the axis and, therefore, the mass $M_1$ per unit length of the cylinder and its radius $R$ are given by
\begin{eqnarray}
p & = & p_c - \pi G \mu_0^2 r^2, \nonumber \\
R & = & \sqrt{\frac {p_c} {\pi G \mu_0^2}},\\
M_1 & = & \frac {p_c} {G \mu_0}. \nonumber
\end{eqnarray}

Is it possible to find the dependance of the cylinder radius on the central pressure in other cases as well? Let us consider, for example, a polytropic EOS, $p = \alpha \mu^\kappa$, with a dimensionless polytropic index $\kappa$. Substituting the EOS into equation \eref{Balance Equation}, we have
\begin{equation}
4 \pi G r \mu^3 = \alpha \kappa r (2-\kappa) \mu^{\kappa -1}
\mu'^2 -\alpha \kappa \mu^\kappa \mu' -\alpha \kappa r
\mu^\kappa \mu''.
\end{equation}
One of the solutions is $\mu = \left( \frac {\pi G (2-\kappa)^2 r^2} {\alpha \kappa (1-\kappa)} \right)^{ \frac {1} {\kappa-2} }$. This, however, is either zero ($\kappa>2$) on the axis or divergent ($\kappa<2$) and thus it is not physically relevant. For $\kappa=2$ we find the non-divergent solution to be the Bessel function of the first kind $J_0(\sqrt{\frac {2 \pi G} {\alpha}}r)$. In this case the radius of the cylinder does not depend on the central pressure.

This peculiar result can be deduced using dimensional analysis as follows. The radius $R$ of a Newtonian polytropic self-gravitating cylinder is described completely by three quantities: $p_c$, $\mu_c$ and $G$. From these quantities only one dimensionless parameter---$\pi = G R^2 \mu_c^2 \;p_c^{-1}$---can be constructed. Thus, the Buckingham $\pi$ theorem \cite{Buckingham} says, that the physical law must take the form $f(\kappa,\pi)=0$, and, assuming uniqueness, we can write $\pi=g^2(\kappa)$, i.e.
\begin{equation}
  R = g(\kappa) \sqrt{p_c \over G \mu_c^2} =
   G^{-{1\over 2}} g(\kappa) {\left( p_c \over \mu_c^\kappa
\right)}^{\!\!\!{1\over \kappa}}{p_c^{\kappa-2\over 2\kappa}}.
\end{equation}
Since ${p_c / \mu_c^\kappa}$ is constant for a given EOS, the derivative
\begin{equation}\label{Derivative radius to central pressure}
  {dR\over dp_c} = {\kappa-2\over 2\kappa} \; {R \over p_c}
\end{equation}
suggests that if a unique solution of hydrostatic equations exists, then the sign of $dR/dp_c$ will be the same as the sign of $\kappa-2$. Moreover, for $\kappa=2$ the radius of the cylinder will not depend on $p_c$ at all.

We note that the same argument cannot be used in the relativistic case since then we have another quantity, namely $c$, the velocity of light, which we need to take into account. In addition to the dimensionless quantity $\pi$ mentioned above, we can also construct the dimensionless central pressure $\Pi_c = p_c / c^2 \mu_c$. Consequently, $f(\kappa,\pi, \Pi_c)=0$, and we do not find the derivative (\ref{Derivative radius to central pressure}) explicitly.

To demonstrate the dependance of these results on the symmetry of the configuration, we give the results for Newtonian spheres. The balance equation reads
\begin{equation}\label{Balance Equation - Sphere}
4 \pi G r \mu^3 = rp'\mu'-2\mu p' -\mu r p''.
\end{equation}
For incompressible fluid of density $\mu_0$, pressure is again a quadratic function of the distance from the center. The total mass $M$ contained within the sphere and its radius $R$ read
\begin{eqnarray}
p & = & p_c - \frac {2} {3} \pi G \mu_0^2 r^2, \nonumber \\
R & = & \sqrt{\frac {3p_c} {2\pi G \mu_0^2}},\\
M & = & \frac {p_c} {G \mu_0^2} \sqrt{\frac {6p_c} {\pi G}}. \nonumber
\end{eqnarray}
The dimensional analysis for a polytropic EOS yields exactly the same results as in the cylindrical case.
\section[]{Conicity outside general perfect-fluid cylinders} \label{The need of conicity outside general perfect-fluid cylinders}

We have seen that the conditions at the axis determine uniquely the fields both inside and outside the cylinders. In particular, we can choose the axis to be regular or to have an inner conicity $\mathcal{C}^* \not= 1$, and the conicity parameter $\mathcal{C}$ of the external field is determined uniquely (cf equation \eref{Conicity Equation}). In section 6, we have shown numerically that the conicity parameter $\mathcal{C}$ outside relativistic self-gravitating cylinders of incompressible fluid is always greater than 1. Is that the case for any cylinder of perfect fluid? Let us investigate what happens if one just puts $\mathcal{C}=1$ in the exterior Levi-Civita metric. We consider the cylinders discussed by Stela and Kramer \cite{Stela and Kramer} and use their coordinate system. They start out from the following metric describing a solid cylinder composed of perfect fluid:
\begin{equation}\label{Kramer}
ds^2 = -e^{2x}dt_-^2 + \frac {yz-1} {8\pi p} dx^2 + e^{-2x}\left( e^{2k} d\xi^2 + e^{2h} d\varphi_-^2 \right),
\end{equation}
where $t_-$, $x$, $\xi$ and $\varphi_- \in [0,2\pi)$ are the temporal, radial, axial and azimuthal coordinates, respectively, $p$ is pressure, and $y, z, k, h$ are functions of $x$. This applies to the range $x \in [0,x_1)$, where $x_1$ denotes the surface of the cylinder, $p(x_1)=0$.

Outside, they use Levi-Civita spacetime in the form
\begin{equation} \label{Coordinate System}
ds^2 = -\rho^{2m} dt_+^2 + \rho^{-2m}
\left[ \rho^{2m^2} (dz^2 + d\rho^2) + \rho^2
d\varphi_+^2 \right],
\end{equation}
i.e., {\it without a general conicity parameter}; so the metric (\ref{Coordinate System}) is our metric (\ref{LC without rho_0}) with $\mathcal{C}=1$. It is claimed in \cite{Stela and Kramer} that these two spacetimes, described by \eref{Kramer} and \eref{Coordinate System}, can be smoothly joined on the surface $p=0$. However, this cannot be achieved if the complete spacetime is to represent a solid cylinder and its external gravitational field with a fixed range of the angular coordinate, $\varphi$, throughout the whole spacetime. We have to consider a general conicity parameter $\mathcal{C} \not= 1$.

To demonstrate this, we shall use Israel's formalism \cite{Israel}. We consider two hypersurfaces located at $x=x_1$ inside and at $\rho=\rho_+$ in the outside spacetime. These two hypersurfaces are intrinsically flat and, therefore, we may identify them locally. Let the coordinates on the resulting hypersurface (representing possibly a matter shell) be chosen as $T \in (-\infty, \infty)$, $Z \in (-\infty,\infty)$ and $\mathit{\Phi} \in [0,2\pi\rho_+^{1-m})$ so that the induced 3-metric is flat:
\begin{eqnarray} \label{Identification}
t_- e^{x_1} & = T = & t_+ \rho_+^m, \nonumber \\
\xi e^{k(x_1)-x_1} & = Z = & z \rho_+^{m(m-1)},\\
\varphi_- e^{h(x_1)-x_1} & = \mathit{\Phi} = & \varphi_+ \rho_+^{1-m}. \nonumber
\end{eqnarray}
The induced surface (3-dimensional) energy-momentum tensor calculated from the jump of the extrinsic curvatures (see \cite{Israel}) then turns out to be
\begin{eqnarray} \label{Induced_Tensor}
8\pi S_{TT} = \sqrt {\frac {8\pi p} {yz-1}} \; (y+z-2) - \rho_+^{m-m^2-1} (1-m)^2, \nonumber\\
8\pi S_{ZZ} = \rho_+^{m-m^2-1} - \sqrt {\frac {8\pi p} {yz-1}} \; z, \\
8\pi S_{\mathit{\Phi}\mathit{\Phi}} = \rho_+^{m-m^2-1} m^2 - \sqrt {\frac {8\pi p} {yz-1}} \; y, \nonumber
\end{eqnarray}
where functions $y$ and $z$ are evaluated on the surface of the cylinder ($p=0$). Einstein's equations yield the following relations for the metric functions
\begin{equation}\label{Einstein Equations}
\begin{array}{rclrcl}
\dot{y} & = & (1-yz) (\mathcal{F}y-2), \hspace{1cm} \dot{z} & = & (1-yz) (\mathcal{F}z-2), \\
0 & = & \dot{p} + (\mu + p),
\end{array}
\end{equation}
where $\mathcal{F}= (\mu +3p)/2p$ (see equations (1), (2) in \cite{Stela and Kramer}). On the surface, $G_{xx}=0$ then implies $yz=1$, and the fraction under the square root symbols in (\ref{Induced_Tensor}) is thus of the `$0/0$' type. Using (\ref{Einstein Equations}) and l'Hospital's rule, we find on the surface
\begin{eqnarray}\label{Surface_Conditions}
\dot{z} = \dot{y}/y^2, \\
\sqrt {\frac {8\pi p} {yz-1}} = \sqrt {-\frac {4\pi \mu y} {\dot{y}}}. \nonumber
\end{eqnarray}
If the inner solution is to be smoothly extended into the Levi-Civita spacetime {\it without} any matter shell on the surface (no `surface layer'), $S_{AB}$ must be zero. Combining equations (\ref{Induced_Tensor}) and (\ref{Surface_Conditions}), we get on the surface
\begin{eqnarray}
\sqrt {-\frac {4\pi \mu y} {\dot{y}}} \; (y+\frac{1}{y}-2) = \rho_+^{m-m^2-1} (1-m)^2, \nonumber\\
\sqrt {-\frac {4\pi \mu y} {\dot{y}}} \; \frac{1}{y} = \rho_+^{m-m^2-1}, \\
\sqrt {-\frac {4\pi \mu y} {\dot{y}}} \; y = \rho_+^{m-m^2-1} m^2. \nonumber
\end{eqnarray}
Comparing the last two equations, we see that $y(x_1)=\pm m=1/z(x_1)$. Then the first equation yields
\begin{equation}\label{Y=1/Z}
y(x_1) = m = 1/z (x_1).
\end{equation}
This leaves us with only one equation fixing the outer radius of the cylinder:
\begin{equation}\label{No-Layer-Condition}
\sqrt {-\frac {4\pi \mu} {\dot{y}(x_1)}} = \sqrt{m} \; \rho_+^{m-m^2-1}.
\end{equation}
If one is only interested in the value of the Levi-Civita parameter $m$ this equation is sufficient. However, there is one more (global) junction condition, namely that the proper length of a hoop with constant $T, Z$ must be the same as measured on both sides of the surface of the cylinder:
\begin{equation}
2\pi e^{h(x_1)-x_1} = 2\pi \rho_+^{1-m},
\end{equation}
so
\begin{equation} \label{Hoop-Condition}
h(x_1) = x_1 + (1-m) \ln \rho_+.
\end{equation}
Integrating Einstein's equations from the axis to the surface where $p(x_1)=0$, we infer the value of $m$ outside just by reading off the value of $y(x_1)=1/z(x_1)$ as seen from equation (\ref{Y=1/Z}). However, to ensure that there be no surface layer at $\rho_+$, we have to require that also conditions (\ref{No-Layer-Condition}) and (\ref{Hoop-Condition}) be fulfilled. There are thus two conditions but only one free parameter---the outer radius $\rho_+$. Combining equations (\ref{No-Layer-Condition}) and (\ref{Hoop-Condition}), we see that any solution must satisfy the relation
\begin{equation}\label{Overall-Condition}
\sqrt {-\frac {4\pi \mu} {\dot{y}}} = \sqrt{y} \; e^{(h-x_1) \frac{y-y^2-1}{1-y}},
\end{equation}
where all quantities are evaluated at the surface $x=x_1$. For a given EOS, the values of all these quantities are determined by a single parameter---the central pressure. This means that, generally, we cannot choose the central pressure arbitrarily since it has to satisfy \eref{Overall-Condition}. Therefore, only special cylinders will comply with this condition. Moreover, we have shown in section 6 that in case of incompressible fluid one gets $\mathcal{C}>1$ for {\it any} central pressure (see figure \ref{Full Cylinders-m and C} and expansion \eref{Series for c}) and thus none of the cylinders composed of incompressible fluid can satisfy (\ref{Overall-Condition}) because we derived this equation starting from the Levi-Civita metric (\ref{Coordinate System}) with $\mathcal{C}=1$.

Surprisingly enough, there exist EOSs for which special cylinders can be constructed that do satisfy this condition and hence the outside Levi-Civita metric with $\mathcal{C}=1$ applies. To give an explicit example, let us consider the Evans analytical solution \cite{Evans}, which is also discussed in \cite{Stela and Kramer}. The equation of state is $\mu = \mu_0 + 5p$; $\mu_0>0$ is (arbitrarily chosen) density at the surface where $p=0$. In the coordinate system of \cite{Stela and Kramer}, the solution can be written as follows:
\begin{eqnarray}
\frac{p(x)}{\mu_0}\equiv \Pi(x)=\frac{1}{6} \left( \frac{a^2}{4} e^{-6x}- 1 \right), \nonumber \\
\Pi_c=\frac{1}{6} \left( \frac{a^2}{4}- 1 \right), \qquad e^{3x_1} = \sqrt{1+6 \Pi_c}, \nonumber \\
y(x) = \frac {a^2-4e^{3x}} {2(a^2-4e^{3x})} \: , \qquad z(x) = \frac {1-4e^{3x}} {2(1-4e^{3x})}, \\
y(x_1)=\frac{1}{z(x_1)}=m = \frac {2\sqrt{1+6\Pi_c}-2} {4\sqrt{1+6\Pi_c}-1} < \frac {1}{2}, \nonumber \\
\lim_{x \rightarrow x_1} \frac {yz-1}{p} = \frac {1} {\mu_0} \frac {18a}{(2a-1)(a-2)}; \nonumber
\end{eqnarray}
$a$ is an arbitrary constant determining the central dimensionless pressure. The condition (\ref{Overall-Condition}) gives a relation\footnote{The complicated nature of expression (\ref{Restriction}) follows from the fact that the requirement of $\mathcal{C}=1$ outside the cylinder is rather artificial.} between the parameter $a$ (or, equivalently, $\Pi_c$) and the surface density $\mu_0$ that has to be satisfied in order that $\mathcal{C}=1$:
\begin{equation}\label{Restriction}
\fl \hspace{1cm} \left[ \mu_0 / (a-2) \right]^{(a-2)^2} = 2^{2(a^2+2a-2)} 3^{-7a^2+a-1} a^{-3a^2} [\pi (2a-1)^3]^{(2a-1)(a+1)}.
\end{equation}
Therefore, given a specific equation of state, which fixes the value of $\mu_0$, the pressure on the axis cannot be chosen arbitrarily if the outside field has $\mathcal{C}=1$. Not all physically admissible cylinders can be sources of the Levi-Civita field with $\mathcal{C}=1$. Only if $p_c$ (i.e. $\Pi_c = p_c/\mu_0$) is chosen so that relation (\ref{Restriction}) is indeed satisfied, the external field has $\mathcal{C}=1$. However, any value $a>2$ gives a positive value of both the central pressure, $\Pi_c$, and surface density, $\mu_0$, and it thus gives a cylinder with $\mathcal{C}=1$ outside.

If we admit $\mathcal{C} \not = 1$ outside, \eref{Hoop-Condition} changes to
\begin{equation}
h(x_1) = x_1 + (1-m) \ln \rho_+ - \ln \mathcal{C},
\end{equation}
and we can smoothly join {\it any} cylinder composed of perfect fluid to the outer Levi-Civita solution. On the other hand, most characteristic quantities are either independent of $\mu_0$ (e.g., the Levi-Civita parameter $m$, mass per unit coordinate or proper length of the cylinder), or they scale with $\mu_0$ ($R_p = r_p(\Pi_c)/\sqrt{\mu_0}$) so that it is possible to use a single cylinder with a given $\Pi_c$ for the description of properties of a whole family of cylinders. Nevertheless, one does not determine the {\it complete} spacetime geometry outside the cylinders without knowing $\mathcal{C}$. If, for example, one studies the focusing of light rays passing on both sides of a fluid cylinder, one needs to know $\mathcal{C}$.

\section*{References}\label{References}
\addtocontents{toc}{\contentsline {section}{\numberline {}\hspace{-0.75cm} References}{\pageref{References}}}


\begin{thebibliography}{99}
 \bibitem {Bicak} Bi\v{c}\'ak J 2000 Selected solutions of Einstein's field equations: their role in general relativity and astrophysics {\it Einstein's Field Equations and Their Physical Implications. Selected Essays in Honour of J\"{u}rgen Ehlers (Lecture Notes in Physics} vol 540{\it )} ed B G Schmidt (Berlin: Springer) pp 1-126
 \bibitem {Bondi} Bondi H 2000 {\it Proc. R. Soc.} A {\bf 456} 2645
 \bibitem {Mena} Mena Marug\'{a}n G A 2000 {\it Phys. Rev.} D {\bf 63} 024005\\
          Manojlovi\v{c} N and Mena Marug\'{a}n G A 2001 \CQG {\bf 18} 2065
 \bibitem {Goncalves} Goncalves S M C V 2002 {\it Phys. Rev.} D {\bf 65} 084045\\
          Goncalves S M C V 2003 \CQG {\bf 20} 37
 \bibitem {Bonnor} Bonnor W B 1999 The static cylinder in general relativity {\it On Einstein's Path: Essays in Honor of Engelbert Schucking} \; ed A Harvey (New York: Springer) pp 113-9
 \bibitem {BZ} Bi\v{c}\'ak J and \v{Z}ofka M 2002 \CQG {\bf 19} 3653
 \bibitem {Exact Solutions} Stephani H, Kramer D, Maccallum M A H, Hoenselaers C and Herlt E 2003 {\it Exact Solutions to Einstein's Field Equations} \; 2nd edn (Cambridge: Cambridge University Press)
 \bibitem {Evans} Evans A B 1977 \JPA {\bf 10} 1303
 \bibitem {Stela and Kramer} Stela J and Kramer D 1990 {\it Acta Phys. Pol.} B {\bf 21} 843
 \bibitem {Carot et al.} Carot J, Senovilla J M M and Vera R 1999 \CQG {\bf 16} 3025
 \bibitem {Kramer-full} Kramer D 1987 \CQG {\bf 5} 393
 \bibitem {Teixeira} Teixeira A F, Wolk I and Som M M 1977 \NC B {\bf 41} 387
 \bibitem {Bronnikov} Bronnikov K A 1979 \JPA {\bf 12} 201
 \bibitem {Haggag} Haggag S and Desokey F 1996 \CQG {\bf 16} 3221
 \bibitem {Scheel} Scheel M A, Shapiro S L and Teukolsky S A 1993 \PR D {\bf 48} 592
 \bibitem {Ehlers1} Ehlers J 1997 \CQG {\bf 14} A119
 \bibitem {Ehlers2} Ehlers J 1998 The Newtonian limit of general relativity {\it Understanding Physics} ed A K Richter (Katlenburg, Lindau: Copernicus Gesellschaft e.V.) pp 1-13
 \bibitem {Philbin} Philbin T G 1996 \CQG {\bf 13} 1217
 \bibitem {Israel} Israel W 1966 \NC B {\bf 44} 1\\
          Israel W 1967 \NC B {\bf 48} 463 (erratum)
 \bibitem {Bonnor-Interpretation} Bonnor W B 1991 {\it Gen. Rel. Grav.} {\bf 24} 551
 \bibitem {VW} Vishveswara C V and Winicour J 1977 \JMP {\bf 18} 1280
 \bibitem {Anderson} Anderson M R 1999 \CQG {\bf 16} 2845
 \bibitem {Thorne} Thorne K S 1965 {\it PhD Thesis} Princeton University, available from University Microfilms Inc., Ann Arbor, MI
 \bibitem {Rendall-Schmidt} Rendall A D and Schmidt B G 1991 \CQG {\bf 8} 985
 \bibitem {Buckingham} Buckingham E 1914 {\it Phys. Rev.} {\bf 4} 345
\end{thebibliography}
\end{document}